\documentclass[twocolumn,showpacs,preprintnumbers,amsmath,amssymb]{revtex4}

\usepackage{ulem} 
\usepackage[usenames]{color}

\usepackage{graphicx}
\usepackage{dcolumn}

\newcommand{\lk}{\left( }
\newcommand{\rk}{\right)}
\newcommand{\ltk}{\left\{ }
\newcommand{\rtk}{ \right\} }
\newcommand{\ldk}{\left[ }
\newcommand{\rdk}{ \right] }

\begin{document}

\title{Brane-induced Skyrmion on $S^3$: baryonic matter in holographic QCD}
 


\author{Kanabu Nawa\footnote{E-mail: nawa@rcnp.osaka-u.ac.jp}}
\affiliation{Research Center for Nuclear Physics (RCNP),
             Osaka University, Mihogaoka 10-1,
             Ibaraki, Osaka 567-0047, Japan}
\author{Hideo Suganuma\footnote{E-mail: suganuma@ruby.scphys.kyoto-u.ac.jp}} 
\affiliation{Department of Physics, Graduate School of Science, 
             Kyoto University, Kitashirakawa, Sakyo, 
             Kyoto 606-8502, Japan}
\author{Toru Kojo\footnote{E-mail: torujj@quark.phy.bnl.gov}}
\affiliation{RIKEN BNL Research Center, Brookhaven National Laboratory,
             Upton, NY 11973, USA}

\begin{abstract}
We study baryonic matter in holographic QCD
with ${\rm D}4/{\rm D}8/\overline{{\rm D}8}$ multi-D brane system
in type IIA superstring theory.
The baryon is described as the ``brane-induced Skyrmion'',
which is a topologically non-trivial chiral soliton
in the four-dimensional meson effective action
induced by holographic QCD.
We employ the ``truncated-resonance model'' approach for the
baryon analysis, including pion and $\rho$ meson fields
below the ultraviolet cutoff scale $M_{\rm KK}\sim 1{\rm GeV}$,
to keep the holographic duality with QCD.
We describe the baryonic matter in large $N_c$
as single brane-induced Skyrmion on the
three-dimensional closed manifold $S^3$ with finite radius $R$.
The interactions between baryons are simulated by
the curvature of the closed manifold $S^3$, and the decrease
of the size of $S^3$ represents the increase of the total
baryon-number density in the medium in this modeling.
We investigate the energy density, the field configuration,
the mass and the root-mean-square radius of
single baryon on $S^3$ as the function of its radius $R$.
We find a new picture of ``pion dominance''
near the critical density in the baryonic matter,
where all the (axial) vector meson fields disappear
and only the pion field survive.
We also find the ``swelling'' phenomena of the baryons as
the precursor of the deconfinement, and propose the mechanism
of the swelling in general context of QCD.
The properties of the deconfinement and
the chiral symmetry restoration in the baryonic matter
are examined by taking the proper order parameters.
We also compare our truncated-resonance model with
another ``instanton'' description of the baryon
in holographic QCD, considering the role of cutoff scale $M_{ {\rm KK} }$.

\end{abstract} 
\maketitle
\section{Introduction}
The concept of ``holography''
is first introduced by D.~Gabor in 1948~\cite{Gab}
as a new technique of the optical physics to playback 
the three-dimensional information
onto the two-dimensional plate as a
hologram.
In 1997,
this concept of holography 
comes to get a new appearance
in the framework of the superstring theory
as the duality between two theories
belonging to the different spatial dimensions.
It is first 
proposed by Maldacena~\cite{Mal}
as the AdS/CFT correspondence between
${\rm AdS}_5\times S^5$ supergravity and 
${\cal N}$=4 
SUSY Yang-Mills theory
through ${\rm D}3$ brane in type IIB superstring theory.
In more general point of view, an 
essential element of the holography
is ``${\rm D}_p$ brane'' as the $(p+1)$-dimensional
membrane in the ten-dimensional space-time.
In fact, ${\rm D}_p$ brane appears as the soliton, i.e.,
the condensed object of the fundamental strings. 
The ${\rm D}_p$ brane
has two important aspects as follows:
$(p+1)$-dimensional gauge theory appears {\it on}
a surface of the ${\rm D}_p$ brane, and 
$\{(p+{\it 1})+1\}$-dimensional supergravity
appears {\it around} the ${\rm D}_p$ brane.
(The italic ``{\it 1}'' denotes the radial dimension 
with non-trivial curvature around
the ${\rm D}_p$ brane, indicating the existence of the gravity).
Actually, the
concept of holography indicates
the duality between the $(p+1)$-dimensional gauge theory 
without the gravity
and $\{(p+{\it 1})+1\}$-dimensional supergravity
mediated by the ${\rm D}_p$ brane,
and the gauge interaction as a ``hologram''
on the surface of the ${\rm D}_p$ brane 
is to give the supergravity as a ``vision'' in the extra-dimension.

One of the most essential properties of the holography is 
the ``strong-weak duality'' between the gauge theory and the
supergravity:
the coupling strengths are transversely related with each other. 
Therefore the holography provides a remarkable possibility that 
non-perturbative aspects of one side can be analyzed by 
the other dual side just with the tree-level calculations.
%
Then, if we find the special configurations of ${\rm D}$ branes
 reflecting QCD on their surfaces,
non-perturbative aspects of QCD
can be successfully examined from the tree-level dual supergravity side.
This is the strategy of the holographic QCD.

There exist several trials to find the special configurations of 
${\rm D}$ branes reflecting QCD.
Eventually,
in 2005, Sakai and Sugimoto succeeded in constructing 
QCD with 
massless quarks and gluons 
from the fluctuation modes of the open strings 
on the D4/D8/$\overline{\rm D8}$ 
multi-D brane configurations in type IIA superstring theory~\cite{SS},
called Sakai-Sugimoto model,
which is one of the most realistic model of holographic QCD.
By using this model, many phenomenological properties of {\it mesons}
belonging to the non-perturbative aspects of QCD like 
meson mass spectra,
hidden local symmetry~\cite{BKY},
vector meson dominance~\cite{Sakurai},
KSRF relation~\cite{KSRF},
GSW model~\cite{GSW}, etc., are 
successfully
derived from the tree-level dual supergravity calculations.
In this sense,
holographic QCD is often regarded as the ``unified meson theory''.
On the other hand,
 baryon description is not straightforward in this approach
 since the classical supergravity is found to be dual with 
 the strong-coupling ``large-$N_c$'' QCD,
 where baryons do not directly appear 
 as dynamical degrees of freedom~\cite{tH}.

In our
previous work,
we gave the first study of the baryon
as a non-trivial topological soliton 
in the four-dimensional meson effective action
derived from holographic QCD.
We call this topological soliton as
a ``brane-induced Skyrmion''~\cite{NSK,NSK1}.
%
%
Especially we 
included
pions and $\rho$ mesons 
appearing below the Kaluza-Klein mass scale 
$M_{\rm KK}\sim 1{\rm GeV}$,
which is often called the ``truncated-resonance model''
for the baryon analysis.
 
Actually, 
$M_{\rm KK}$ plays the role as the
ultraviolet cutoff scale of the holographic approach.
In fact,
there appear infinite number of ``non-QCD modes''
with mass scale $\sim O(M_{\rm KK})$ in holographic QCD
like gluinos and Kaluza-Klein modes.
In this sense,
the duality with QCD could be maintained below $M_{\rm KK}$ 
as the ultraviolet cutoff.

The
appearance of certain cutoff scale $M_{\rm KK}$ should be essential
for the holographic approach to be dual 
of realistic QCD with 
confinement and chiral
symmetry breaking as the non-SUSY natures.
In the holographic model
with ${\rm D}4/{\rm D}8/\overline{{\rm D}8}$ multi-D brane system,
D4 branes are $S_1$-compactifiled with $M_{\rm KK}$ scale,
to give the complete SUSY breaking and its resulting non-conformal natures
of QCD like finite string tension and chiral condensate.
These considerations suggest that $M_{\rm KK}$ should be so much respected,
giving our truncation of meson resonances at $M_{\rm KK}$ for baryon analysis
(See, Sec.~\ref{MEA} for details).
%
%
%

%
%
%

Recently a baryon is also described as an ``instanton''
on the five-dimensional gauge theory of D8 branes with 
D4 supergravity background~\cite{HSSY,HRYY,HSS,HMY,d_Inst}.
The instanton is introduced {\it before} the mode expansion of the 
five-dimensional gauge field into mesons,
so that the baryon as the instanton is to be composed by the 
infinite number of color-singlet modes with mesonic quantum number 
even above $M_{\rm KK}$ scale.
However, we consider that such color-singlet modes above  $M_{\rm KK}$ 
might not directly correspond to physical mesons in QCD,
because the duality with QCD is mentained below  $M_{\rm KK}$.
Furthermore, there also exist infinite number of other non-QCD modes
above $M_{\rm KK}$, which could also affect the baryon properties if they were included.
Therefore, in contrast to instanton models,
we severely respect the cutoff scale $M_{\rm KK}$ and
truncate the meson resonances at  $M_{\rm KK}$,
to keep the duality with QCD.
More comprehensive discussions with instantons 
are summarized in Sec.~\ref{SO} .

In present work,
we newly consider the extension of the holographic model to dense QCD.
Due to the non-abelian nature of QCD,
various realizations
are expected
in the vacuum itself with finite temperature and density,
called ``QCD phase diagram''.
Up to now,
interesting phase structures are proposed by using some low-energy
effective theories of QCD, e.g., confined phase with 
mesons and baryons,
deconfined phase with quark-gluon plasma (QGP), 
chiral symmetry broken phase with mass generation~\cite{NJL1}, color
superconductivity
as diquark condensation~\cite{BL, RW, ARW, RSSV, NNY}, etc.
In fact, some wisdom about QGP gives the insight
for the early universe just after the Big Bang~\cite{Uni1}.
Furthermore, the possible QCD phase transitions
in the core region of neutron stars
could affect a lot of their macroscopic features
like moment of inertia, angle velocity, and breaking index~\cite{Nstar}.
%
There also exist several experimental projects
to search the QGP 
in the ultra-relativistic heavy ion collisions 
in RHIC (Relativistic Heavy Ion Collider) at BNL, and 
LHC (Large Hadron Collider) at CERN.
There will also appear 
relatively low-energy 
collision experiments
to make the 
low-temperature high-density object 
in FAIR (Facility for Antiproton and Ion Research) at GSI,
giving some knowledge
about 
the core region of the compact stars.
%
With these background, 
it should be urgently important to
make clear the structure of QCD phase diagram more explicitly
from QCD itself 
with the rich help of experimental data,
which will 
eventually bring about the fundamental understanding of our whole nature.

%
There exists the lattice QCD numerical study as the
first principle calculation of the strong interaction.
However, because of the ``sign problem'',
its applicability is severely restricted near the  
zero-density at finite-temperature regime of the wide QCD phase diagram
 (For some review, see, Ref.~\cite{Ste1}).
Therefore, 
if one succeeds in 
the extension of holographic approach 
to the dense regime,
it should give a new analytical tool for non-perturbative aspects
of the finite density QCD, where
the holography provides the duality
between the strong-coupling gauge theory
and the weak-coupling supergravity.
This is the main aim of our study.  

In this work, 
 we consider the baryonic matter in holographic QCD
 as the extension of the holographic approach to dense QCD.
 Especially,
we treat the baryonic matter with large-$N_c$
because the holographic QCD is derived as a large-$N_c$ effective
theory.
As the general property of large-$N_c$ QCD~\cite{tH},
the static baryon mass is proportional to $O(N_c)$,
so that its kinetic energy becomes $O(N_c^{-1})$.
There also exist the quantum effects like the zero point quantum fluctuation energy $E_0$
and also the baryon mass splitting $\Delta m$ within the baryonic matter,
while these correspond to the higher order contributions of the $1/N_c$ expansions
as $E_0\sim O(N_c^{0})$ and $\Delta m\sim O(N_c^{-1})$~\cite{ANW}.
Such large-$N_c$ countings indicate that, for sufficiently large $N_c$,
the kinetic energy and the quantum effects within the baryonic matter
can be suppressed relative to the static mass, and
the baryonic matter comes into the static Skyrme matter.
Such static Skyrme matter was first
analyzed by Klebanov~\cite{Kle},
by placing the Skyrme soliton solutions
periodically along the three-dimensional cubic lattice,
which would correspond to the nuclear crystal
in the deeper interior of neutron stars.

Since the cubic lattice treatment is rather cumbersome,
we employ a mathematical trick to analyze
such static Skyrme matter, proposed by Manton and Ruback~\cite{MR}.
In order to represent 
some high-density state of the 
multi-Skyrmion system 
on the three-dimensional flat space 
${\boldmath \mbox{${\rm R}$}}^3$,
single Skyrmion is alternately placed on a surface of the three-dimensional
closed manifold $S^3$ with a finite radius.
In fact, the multi-Skyrmion system on 
${\boldmath \mbox{${\rm R}$}}^3$
and the single Skyrmion on $S^3$
can be related with each other
through the compactification of the boundary
of a unit cell on 
${\boldmath \mbox{${\rm R}$}}^3$
shared by one Skyrmion.
The interactions between baryons on 
${\boldmath \mbox{${\rm R}$}}^3$
are simulated by the curvature of the manifold $S^3$,
and 
decreasing the radius of $S^3$ represents
the increase of the 
baryon-number density 
in the medium in this modeling.
Actually, by taking such mathematical simplification,
one could avoid some complicated 
analysis like Monte Carlo simulations on the
three-dimensional
cubic lattice~\cite{Kle}, and he can
get some physical intuitions
as for the baryonic matter qualitatively and even quantitatively~\cite{MR}.
Therefore, by placing the single brane-induced Skyrmion
on the closed manifold $S^3$,
we try to analyze the typical features of baryonic matters
in holographic QCD. Especially, in this analysis,
the roles of $\rho$ mesons in the dense baryonic matter will 
be examined in detail
from the holographic point of view.

There exist a lot of works about the extension of the holographic
approaches
to dense SUSY QCD (See, Ref.~\cite{d_SUSYQCD} and references therein).
%
With the bottom-up construction of the AdS/QCD models~\cite{Hill1, Braga1,Son,Hill2, Son2, Brod1, Brod2,Poma1},
it was applied to dense QCD by introducing the bi-nucleon
condensate~\cite{HW}.
After the discovery of the Sakai-Sugimoto model~\cite{SS} as one of the most
reliable holographic top-down approaches to non-SUSY QCD with massless
quark flavors,
there exist several proposals about the extension of this model to dense QCD.
For example, in Refs.~\cite{Kim, Hori},
the baryon chemical potential $\mu_B$ is introduced by the asymptotic value of
a ${\rm U}(1)$ gauge field on the D8-$\overline{{\rm D}8}$ branes
as ${\cal V}_0=-i\mu_B/N_c$,
similarly to the introduction of the chemical potentials
to the chiral perturbation theory
by promoting the global chiral symmetry 
to local gauge one~\cite{Chemi1,Chemi2}.
In these holographic analyses, 
the density dependence of the ``local'' properties of
mesons and baryons like their masses and coupling constants, and also
the phase structure of QCD have been successfully discussed.
Now, in our paper,
we treat the baryon as a ``non-local'' solitonic object 
as the Skyrmion in the Sakai-Sugimoto model, and
we analyze the dense QCD by the Skyrme matter.
Therefore,
adding to the informations about the local natures of hadrons and QCD
phase structure, we can extensively examine the internal structure of
the baryon like the size and the field configurations.
%
%

Here we show the organization of this paper and its brief summary.
%
In Sec.~\ref{MEA},
we overview the holographic derivation of
the four-dimensional meson effective action
with the pion and $\rho$ meson fields,
reemphasizing the role of
cutoff scale $M_{\rm KK}$
in the holographic framework to be dual of QCD. 
In Sec.~\ref{BISrs},
we analyze the properties of baryons and baryonic matter
in holographic QCD.
In Sec.~\ref{BIS_R3},
we introduce the concept of the 
brane-induced Skyrmion on the three-dimensional flat space 
${\boldmath \mbox{${\rm R}$}}^3$.
Then, in Sec.~\ref{BIS_S3},
we describes the baryonic matter as
the system of single 
brane-induced Skyrmion
on the three-dimensional closed manifold $S^3$.
We derive the expression of
the hedgehog mass and 
Euler-Lagrange equations
for the pion and $\rho$ meson fields as the 
brane-induced Skyrmion
on $S^3$.
Sec.~\ref{NR}
is devoted to
the numerical results and their physical interpretations 
about the baryon nature in dense QCD.
In Sec.~\ref{ECpro},
the  
baryon-number density dependence of the energy density
and field configuration profiles of single baryon
are discussed.
In Sec.~\ref{PDNC},
we  
propose a new striking picture of the ``pion dominance''
near the critical density,
i.e.,
all the (axial) vector meson fields disappear and only pion field survives.
%
In Sec.~\ref{MP},
the baryon-number density dependence of the mass and root-mean-square mass radius
of single baryon are analyzed.
We find some non-linear increase in the size of the baryon
near the critical density as a ``swelling'' phenomena.
In Sec.~\ref{SMPI}, we  
explain the swelling mechanism in the general context of QCD,
and consider its effects on
the stability of $N$-$\Delta$ mixed matter.
In Sec.~\ref{CRDPT},
we examine the features of 
the delocalization phase transitions and
the chiral symmetry restoration 
by choosing proper order parameters,
through which
the relations 
between deconfinement and chiral symmetry restoration
are reconsidered.
In Sec.~\ref{CDT},
we calculate the critical densities of the phase transitions
in the physical units with the experimental inputs
for the pion decay constant $f_\pi(=92.4{\rm MeV})$ 
and $\rho$ meson mass $m_\rho(=776.0{\rm MeV})$.
We find the critical density $\rho_B\simeq 7\rho_0$
in the holographic approach.
Through all of the sections,
by comparing the brane-induced Skyrmion
and standard Skyrmion
without $\rho$ meson fields,
the roles of the vector mesons in the dense baryonic matter
are examined from the holographic point of view.
Sec.~\ref{SO} is devoted to summary and outlook.
In this final section, we compare our truncated-resonance
approach with another ``instanton'' description of
baryons \cite{HSSY,HRYY,HSS,HMY,d_Inst} in the holographic QCD,
paying attention on the role of cutoff scale $M_{ {\rm KK}}$.

\section{Meson effective theory from holographic QCD \label{MEA}}
In this section, we 
overview the derivation of
the four-dimensional meson effective 
action from holographic QCD with ${\rm D}4/{\rm D}8/\overline{{\rm D}8}$
multi-${\rm D}$ brane system in type IIA superstring theory,
called the Sakai-Sugimoto model~\cite{SS}.
The meson effective action is derived without small amplitude
expansion 
to discuss a baryon as a large amplitude chiral soliton.
(For more comprehensive derivations, see our previous paper~\cite{NSK}).
Here we especially emphasize the roles of cutoff scale $M_{\rm KK}$ in
holographic QCD.

First, we review the construction of 
${\rm D}4/{\rm D}8/\overline{{\rm D}8}$ multi-D brane system
and also the supergravity description of D4 branes as the holographic
dual of QCD. 
As the first step, $N_c$ sheets of D4 branes are prepared to 
construct the gluon sector of QCD.
The D4 branes are $S^1$-compactified along one extra-dimension
with a radius as the inverse of the Kaluza-Klein mass scale $M_{\rm KK}$.
There appear ten independent fluctuation modes from the open strings
on the surface of D4 branes, i.e.,
gauge fields ${\cal A}_{\mu=0\sim 3}$, scalar fields $A_4$ and
$\Phi_{i=5\sim 9}$, and also their superpartners as fermions.
By imposing the anti-periodic boundary conditions for all the fermions
along the $S^1$-compactified direction, 
they acquire large masses $\sim O(M_{\rm KK})$.
Then supersymmetry (SUSY) is completely broken and,
due to the radiative corrections, 
all the scalar fields $A_4$ and $\Phi_{i}$ also 
get large masses $\sim O(M_{\rm KK})$.
Because of the $S^1$-compactification,
there also appear the infinite number of the Kaluza-Klein modes with masses 
$\sim O(M_{\rm KK})$.
Therefore, below the $M_{\rm KK}$ scale,
only massless gauge fileds ${\cal A}_{\mu}$ appears.
In this sense, the system of $N_c$ sheets of D4 branes with
the $S^1$-compactification can be viewed as the ${\rm U}(N_c)$
Yang-Mills theory below $M_{\rm KK}$ scale, corresponding to the pure gauge sector of QCD.
As the next step, $N_f$ sheets of D8 and $\overline{{\rm D}8}$ branes are
added to introduce the massless quark flavors of QCD.
$\overline{{\rm D}8}$ has opposite chirality relative to 
D8, providing ${\rm U}(N_f)_L\times {\rm U}(N_f)_R$ chiral
symmetry in this model.
From the fluctuation modes of open strings between
D4 and D8 ($\overline{{\rm D}8}$), there appear massless chiral fermions
as quarks in QCD.
As a whole, massless QCD appears as the ``hologram''
on the surface of ${\rm D}4/{\rm D}8/\overline{{\rm D}8}$ branes.

Then we shift into the gravitational description of D branes from the
extra-dimensions.
The D brane is originally introduced as the fixed edges of open strings
with Dirichlet boundaries.
This also indicates that, through the ``open-closed duality'' for the 
fundamental strings, 
the D brane can also be regarded as the source of closed strings,
giving the graviton in the extra-dimensions outside of the D brane.
In this sense, the D brane can be identified as a highly gravitational
system, i.e., the ``black brane'', allowing the gravitational
description from the extra-dimensions.
In fact, the mass of the D branes is proportional to its sheets
number,
so that, by assuming $N_c\gg N_f$, 
only D4 branes can be represented by the gravitational background and
D8 ($\overline{{\rm D}8}$) branes are introduced as the probes called
``probe approximation'', which corresponds to the quenched approximation
in lattice QCD study~\cite{Rothe}.
Especially the classical supergravity description of D4 branes
is tractable, followed by the local approximation of the strings and
also the suppressions of the string loop effect.
These conditions in the gravitational side around the D branes
give the constraints for the gauge theory side as the QCD on the surface
of D branes as 
\begin{eqnarray}
g_{\rm YM}^4 \ll \frac{1}{g_{\rm YM}^2 N_c}\ll 1, \label{Cond1}
\end{eqnarray}
which is achieved by $g_{\rm YM}\rightarrow 0$, $N_c\rightarrow \infty$, 
and 'tHooft coupling: $\lambda\equiv g_{\rm YM}^2 N_c$ fixed and large.
In this sense,
the strong-coupling large-$N_c$ QCD is found to be dual with
the classical supergravity of D4 branes 
with probe D8 ($\overline{{\rm D}8}$) branes,
which is one of the realization of the strong-weak duality 
between the gauge theory and gravitational theory through the
holography.
Therefore, by analyzing the effective action of D8 branes
with D4 supergravity background around the D branes, 
one can analyze the nonperturbative aspects of QCD on the surface of D
branes.

Then we start formal discussions from
the $N_f=2$ non-Abelian Dirac-Born-Infeld (DBI)
action of probe ${\rm D}8$ brane with ${\rm D}4$ supergravity 
background as a probe approximation.
After 
dimensional reductions,
the nine-dimensional DBI action of the probe ${\rm D}8$ brane
with the ${\rm D}4$ supergravity background
can be reduced into a five-dimensional Yang-Mills theory,
belonging to the flat four-dimensional Euclidean space-time $x$, i.e., 
$x_{0\sim 3}$ and the other fifth dimension $z$ with curved measures as 
follows~\cite{SS}:
\begin{eqnarray}
S_{{\rm D}8}^{\rm DBI}-S_{{\rm D}8}^{\rm DBI}|_{A_M\rightarrow 0}&=&
\kappa\int d^4 x dz\mbox{tr}\{
\frac{1}{2}K(z)^{-1/3}F_{\mu\nu}F_{\mu\nu}\nonumber\\
&&+K(z) F_{\mu z}F_{\mu z}\}+O(F^4),
\label{5dimDBI_2}
\end{eqnarray}
where $A_M$ is the gauge field and 
$F_{MN}=\partial_M A_N-\partial_N A_M+i\ldk A_M,A_N\rdk $ 
($M, N=0\sim 3, z$) 
is the field strength tensor in five-dimensional space-time 
$(x_{0\sim 3}, z)$ of the probe ${\rm D}8$ branes.
In the action (\ref{5dimDBI_2}),
$M_{\rm KK}=1$ unit is taken,
and 
the overall factor $\kappa$ is defined as
\begin{eqnarray}
\kappa\equiv\frac{\lambda N_c}{216\pi^3}.\label{kappa_def}
\end{eqnarray}
(Note here that
 we use the value of $\kappa$ as a half of that 
 in Ref.~\cite{NSK}, 
 taking away the misleading factor $2$ in Eq.(16) of Ref.~\cite{NSK}.
 All the formula and numerical results can be scaled
 by the factor $\kappa$,
 so that the discussions in Ref.~\cite{NSK} are not altered.) 
The functional $K(z)\equiv 1+z^2$ in the action (\ref{5dimDBI_2}) expresses the nontrivial curvature in the extra fifth dimension $z$ induced by the supergravity 
background of the D4 brane.
The gravitational energy of ${\rm D}8$ brane, i.e.,
$S_{{\rm D}8}^{\rm DBI}|_{A_M\rightarrow 0}$ is subtracted 
in the action (\ref{5dimDBI_2}) 
as the vacuum relative to the gauge sectors.

In the ${\rm D}4/{\rm D}8/\overline{{\rm D}8}$
multi-${\rm D}$ brane configurations,
color quantum number is carried only by the $N_c$-folded ${\rm D}4$
branes.
Therefore, after the supergravity description of ${\rm D}4$ branes,
there is no colored particles from the fluctuation modes of
open strings on the residual probe ${\rm D}8$ branes.
This is regarded as some holographic manifestation of ``color confinement''
in low-energy scale of QCD.
In fact, gauge field $A_{M=0\sim 3, z}$
in the action (\ref{5dimDBI_2})
is color-singlet and obeys the adjoint representation of $U(N_f)$ group,
eventually producing the meson degrees of freedom after some proper mode
expansions in the holographic QCD.

In this paper, we treat the 
non-trivial leading order 
of $1/N_c$ and $1/\lambda$
expansions in the holographic QCD
as the five-dimensional Yang-Mills 
action (\ref{5dimDBI_2}) with $O(F^2)$.
In general, there also exists the Chern-Simons
(CS) term to avoid anomalies in the superstring theory.
By introducing $\omega$ meson degrees of freedom
as the $U(1)$ sector of $\rho$ meson fields 
in the holographic approach,
$\omega$ meson indirectly couples with pions via (axial) vector mesons
after the proper mode expansions of CS term,
which is regarded as a general representation of the 
Gell-Mann - Sharp - Wagner
(GSW) model~\cite{GSW}.
The roles of $\omega$ mesons for low-energy meson dynamics 
and also chiral solitons have been
traditionally examined in some of QCD phenomenologies~\cite{KFLiu}.
Actually, however, the CS term is found to be $O(\lambda^0)$,
i.e., the higher order contributions of 'tHooft coupling expansion 
relative to the $O(F^2)$ of the Yang-Mills action (\ref{5dimDBI_2}) 
with $O(\lambda^1)$, which is manifestly shown in holographic QCD.
Furthermore,
most part of CS term includes one time-derivative of pion fields
and does not affect the static properties of hedgehog solitons.
Therefore
we neglect the CS term 
along the discussions below
for the argument of the non-perturbative (strong coupling) properties of
QCD.

In the holographic approach,
the pion field is introduced as the 
Wilson line of the fifth gauge field $A_z$, i.e.,
a path-ordered product of the fifth gauge field 
along the $z$ direction~\cite{SS, NSK, Son}
as
\begin{eqnarray}
U(x^{\mu})=P \exp\ltk
 -i\int_{-\infty}^{\infty}dz'A_z(x_\mu,z')\rtk \in U(N_f).
\label{pion1}
\end{eqnarray}
One can also introduce the variables $\xi_\pm(x_\mu)$ as
\begin{eqnarray}
\xi_{\pm}^{-1}(x^{\mu})=P\exp\ltk -i\int_{z_0(x_\mu)}^{\pm\infty}
dz'A_z(x_\mu,z')\rtk \in U(N_f),\nonumber\\
\label{xi1}
\end{eqnarray}
where $z_0(x_\mu)$ is a 
single-valued arbitrary function of $x_\mu$.
Then the pion field (\ref{pion1}) can be written as
\begin{eqnarray}
U(x^{\mu})=\xi_{+}^{-1}(x_\mu)\xi_-(x_\mu),\label{pion2}
\end{eqnarray}
some resemble formula of which can also be found in 
the traditional approach of hidden local symmetry~\cite{BKY}.

Now we take ``$A_z=0$ gauge'' and also 
``$\xi_{+}^{-1}(x_\mu)=\xi_-(x_\mu)(\equiv \xi(x_\mu))$ gauge'' for the $U(N_f)$ 
gauge symmetry in the action (\ref{5dimDBI_2}) of the probe ${\rm D}8$
brane.
The $A_z=0$ gauge is similar to the unitary gauge in the non-Abelian
Higgs theory;
fifth gauge field $A_z$
performs as a scalar field in four-dimensional space-time 
$x_{\mu=0\sim 3}$, and it is eaten by the four dimensional
gauge field $A_\mu$ to give the mass generation of gauge field,
especially the (axial) vector mesons as a part of $A_\mu$. 
In this sense the masses of the (axial) vector mesons
come from the Higgs mechanism in five-dimensional space-time
with the $U(N_f)$ gauge symmetry breaking
on the probe ${\rm D}8$ brane.
The $\xi_{+}^{-1}(x_\mu)=\xi_-(x_\mu)$ gauge is also essential
to get the low-energy effective theory of QCD
with proper parity and G-parity classification
in a manifest way~\cite{NSK}.
With these gauge fixings,
the five-dimensional gauge field $A_\mu(x_N)$ can be mode-expanded
into the four-dimensional parity and G-parity eigen states with
proper complete orthogonal basis $\psi_{\pm}(z)$ and $\psi_n(z)$
($n=1,2,\cdots$) as follows~\cite{SS, NSK}:
\begin{eqnarray}
A_\mu(x_N) 
&=& l_\mu(x_\nu)\psi_+(z)+
           r_\mu(x_\nu)\psi_-(z)\nonumber\\
&&+\sum_{n\geq 1}B_\mu^{(n)}(x_\nu)\psi_n(z),\label{limit2_lr}\\
 l_\mu(x_\nu) &\equiv& \frac{1}{i}\xi^{-1}(x_\nu)\partial_\mu\xi(x_\nu), \label{l_def}\\
     r_\mu(x_\nu) &\equiv &\frac{1}{i}\xi (x_\nu)\partial_\mu\xi^{-1}(x_\nu),\label{r_def}
\end{eqnarray} 
where $l_\mu$ and $r_\mu$ are left and 
right currents of pion fields, respectively.
The basis $\psi_{\pm}(z)$ are introduced 
to support whole of the gauge field $A_\mu(x_N)$
at the boundary $z\rightarrow \pm\infty$ as
$\psi_\pm(z\rightarrow \pm\infty)=1$ and
$\psi_\pm(z\rightarrow \mp\infty)=0$ as 
\begin{eqnarray}
\psi_\pm(z)&\equiv& \frac{1}{2}\pm\hat{\psi}_0(z),\label{hat_psi_0}\\
\hat{\psi}_0(z)&\equiv&
 \frac{1}{\pi}\arctan z.\label{psi_n}
\end{eqnarray}
In order to diagonalize the five-dimensional Yang-Mills action (\ref{5dimDBI_2})with the induced measures $K(z)^{-1/3}$ and $K(z)$ in the fifth dimension $z$, 
the basis $\psi_n$ $(n=1,2,\cdots)$ are taken to be the normalizable eigen-function satisfying
\begin{eqnarray}
-K(z)^{1/3}\frac{d}{dz}\ltk K(z) \frac{d\psi_n}{dz}\rtk =
 \lambda_n\psi_n,
\hspace{4mm}\mbox{($\lambda_1 < \lambda_2 < \cdots$)}\nonumber\\
\label{eigen1}
\end{eqnarray}
with normalization condition as 
\begin{eqnarray}
\kappa\int dz K(z)^{-1/3}\psi_m \psi_n =\delta_{nm}.\label{normal_psi_n}
\end{eqnarray}

In the holographic model, the fields $B_\mu^{(n=1,2,\cdots)}$ in
the mode expansion (\ref{limit2_lr}) are regarded 
as (axial) vector mesons,
belonging to the adjoint representation of the $U(N_f)$ 
gauge group as
$B_\mu=B_\mu^aT^a$.
By substituting the expansion (\ref{limit2_lr}) into the action 
(\ref{5dimDBI_2}),
the mass of $B_\mu^{(n)}$ field is found
with the eigenvalue of oscillating fifth basis 
in Eq.(\ref{eigen1}) as $m_n^2\equiv \lambda_n$, indicating that the origin of meson mass
is the oscillation of meson wave function in the extra fifth dimension.
Furthermore, $A_\mu({x_N})$ is the five-dimensional vector
and $\psi_n(z)$ are the parity eigen state in the $z$ direction
as $\psi_n(-z)=(-)^{n-1}\psi_n(z)$.
Therefore, from the mode expansion (\ref{limit2_lr}),
$B_\mu^{(n)}$ fields have four-dimensional parity transformation as
$B_\mu^{(n)}(-x_\nu)\rightarrow(-)^nB_\mu^{(n)}(x_\nu)$.
These consideration indicates that
vector and axial vector mesons appear alternately in the 
excitation spectra about index $n$ as 
$B_\mu^{(1)}\equiv \rho_\mu,
B_\mu^{(2)}\equiv a_{1\mu},
B_\mu^{(3)}\equiv \rho'_\mu,
B_\mu^{(4)}\equiv a'_{1\mu},
B_\mu^{(5)}\equiv \rho''_\mu,\cdots$.

In our study, we 
construct the four-dimensional meson effective action only with 
pion field $U(x_\nu)$ and $\rho$ meson field 
$B_\mu^{(1)}(x_\nu)\equiv \rho_\mu(x_\nu)$ 
below the Kaluza-Klein mass scale $M_{\rm KK}\sim 1{\rm GeV}$.
Recall that
there appear infinite number of non-QCD modes
with large mass $\sim O(M_{\rm KK})$ in holographic QCD, e.g.,
scalar fields, gluino, and also the Kaluza-Klein modes,
discussed in the first part of this section.
Therefore, the ${\rm D}4/{\rm D}8/\overline{{\rm D}8}$
multi-D brane system can be viewed as QCD
as far as low energy phenomenology below $\sim M_{ {\rm KK}}$
is concerned.
In this sense,
$M_{\rm KK}$ plays the roles as the ultraviolet cutoff scale of the
theory,
so that we include the meson degrees of freedom below $M_{\rm KK}$
for the baryon analysis in later sections,
called ``truncated-resonance model''.

Actually, the appearance of $M_{\rm KK}$ scale with finite value
seems to be essential in the recent holographic analysis.
In the framework of AdS/CFT correspondence without Kaluza-Klein
compactification,
the ${\cal N}=4$ SUSY 
and its resulting conformal symmetry
protect the emergence of the dimensional quantities like
string tension and chiral condensate at the ground state,
because of the cancellation of the radiative corrections between bosons and fermions.
In this sense, SUSY breaking is at lest needed to be dual of QCD
with confinement and chiral symmetry breaking as its vacuum nature.
In the Sakai-Sugimoto model,
$N_c$ sheets of D4 branes are Kaluza-Klein compactified with radius
$M_{\rm KK}^{-1}$, and the SUSY is completely broken
by the field boundary condition along the compactified direction.
Therefore confinement and chiral symmetry breaking {\it could} occur
as the non-SUSY gauge theory.
In fact, the compactified D4 brane is ``non-BPS'', having a ``horizon''
in the supergravity description.
In the holographic framework,
confinement and chiral symmetry breaking {\it do} occur 
on this horizon, giving the lost of ``colored'' information,
and also the geometrical connection of D8 and
$\overline{{\rm D}8}$ branes.
As a whole, the appearance of $M_{\rm KK}$ would be essential 
for the holographic model to be dual
of realistic QCD.

Now, one may regret about the finiteness of $M_{\rm KK}$
as almost 1GeV, which is comparable with the QCD mass scale
$\Lambda_{\rm QCD}$.
In the holographic approach, the QCD mass scale is
introduced by the experimental inputs for the pion decay constant
and $\rho$ meson mass as $f_\pi=92.4{\rm MeV}$ and $m_\rho=776.0{\rm
MeV}$.
These experimental values come from our hadronic world with $N_c=3$
and large but finite 'tHooft coupling $\lambda$,
which may infringe the condition (\ref{Cond1}) to give the effects 
of string length and string loops in the gravitational side.
Therefore, too large $M_{\rm KK}$ can not be taken to neglect
the internal structure of strings on the surface of D4 branes
compactified with radius $M_{\rm KK}^{-1}$, which might eventually gives
the scale $M_{\rm KK} \sim 1{\rm GeV}$.
These considerations indicate that,
by including the effects of string length and loops in the gravitational
side,
$M_{\rm KK}$ would be taken sufficiently large relative to 
$\Lambda_{\rm QCD}$ with fixed $f_\pi$ and $m_\rho$.
Anyway, $M_{\rm KK} \sim 1{\rm GeV}$ essentially appears as the
ultraviolet cutoff scale in the system of probe D8 brane with D4
``classical'' supergravity background to be dual of QCD with proper
dimensional quantities.

By neglecting the higher mass excitation modes of (axial) vector mesons
rather than $\rho$ meson sector with $n=1$ 
in the expansions (\ref{limit2_lr}),
the five-dimensional gauge field $A_\mu(x_N)$
can be written as
\begin{eqnarray}
A_\mu(x_N) 
=l_\mu(x_\nu)\psi_+(z)+
   r_\mu(x_\nu)\psi_-(z)+
 \rho_\mu(x_\nu)\psi_1(z).\nonumber\\
\label{exp_rhomode}
\end{eqnarray}
By taking this mode expansion (\ref{exp_rhomode}) with 
the $A_z=0$ gauge,
five-dimensional field strength
$F_{\mu\nu}$ and $F_{z\mu }$ can be written as
\begin{widetext}
\begin{eqnarray}
F_{\mu\nu}\!\!&=&\!\!\partial_\mu A_\nu-\partial_\nu A_\mu+
             i\ldk A_\mu,A_\nu\rdk\nonumber\\ 
          &=&\!\!\lk\partial_\mu l_\nu-\partial_\nu l_\mu \rk \psi_+ 
           +\lk\partial_\mu r_\nu-\partial_\nu r_\mu \rk \psi_-
          +\lk\partial_\mu \rho_\nu-\partial_\nu \rho_\mu \rk
           \psi_1
              +i\ltk [l_\mu,l_\nu ]\psi_+^2+
                [r_\mu,r_\nu ]\psi_-^2+
                [\rho_\mu,\rho_\nu ]\psi_1^2
           \rtk\nonumber\\
         &&+i\{ \lk [l_\mu,r_\nu ]+[r_\mu,l_\nu ] \rk\psi_+\psi_- 
                 +\lk [l_\mu,\rho_\nu ]+[\rho_\mu,l_\nu ] \rk\psi_+\psi_1    
                 +\lk [r_\mu,\rho_\nu ]+[\rho_\mu,r_\nu ]\rk\psi_-\psi_1 
             \}\nonumber\\
         &=&\!\!-i[\alpha_\mu,\alpha_\nu]\psi_+\psi_- +
             \lk \partial_\mu\rho_\nu -\partial_\nu \rho_\mu \rk \psi_1
             +i[\rho_\mu,\rho_\nu]\psi_1^2
             +i\{\lk[\alpha_\mu,\rho_\nu]+[\rho_\mu,\alpha_\nu]\rk\hat{\psi}_0\psi_1\nonumber\\ 
	       &&\hspace{-1.5mm}+\lk[\beta_\mu,\rho_\nu]+[\rho_\mu,\beta_\nu]\rk\psi_1
              \},\label{fieldst1}\\
F_{z\mu}\!\! &=&\!\!\partial_z A_\mu 
              =\alpha_\mu\partial_z\hat{\psi}_0+\rho_\mu\partial_z\psi_1,\label{fieldst2}               
\end{eqnarray}
\end{widetext}
with axial vector current $\alpha_\mu$ and vector current $\beta_\mu$ of 
pion field as 
\begin{eqnarray}
\alpha_\mu(x_\nu) &\equiv&  l_\mu(x_\nu)-r_\mu(x_\nu), \label{alpha_def}\\
\beta_\mu(x_\nu)  &\equiv&  \frac{1}{2}\ltk l_\mu(x_\nu)+r_\mu(x_\nu)\rtk. \label{beta_def}
\end{eqnarray}
In the derivation of (\ref{fieldst1}) and (\ref{fieldst2}), 
we have used the Maurer-Cartan equations, 
$\partial_\mu l_\nu-\partial_\nu l_\mu+i[l_\mu,l_\nu]=0$ and 
$\partial_\mu r_\nu-\partial_\nu r_\mu+i[r_\mu,r_\nu]=0$.
By substituting Eqs.(\ref{fieldst1}) and (\ref{fieldst2})
 into the five-dimensional Yang-Mills action (\ref{5dimDBI_2})
with $O(F^2)$, we eventually get the
four-dimensional Euclidean meson effective action 
with pions and $\rho$ mesons from holographic QCD
as follows
(derivations in more detail can be found in our previous paper~\cite{NSK}):
\begin{eqnarray}
S_{\rm eff}&\equiv& %
S_{{\rm D}8}^{\rm DBI}-S_{{\rm D}8}^{\rm DBI}|_{A_M\rightarrow 0} \nonumber\\
&=& \kappa\int d^4 x dz \mbox{tr}\ltk
               \frac{1}{2}K(z)^{-1/3}F_{\mu\nu}F_{\mu\nu}+%
               K(z) F_{\mu z}F_{\mu z}\rtk \nonumber\\
                                                                                   \label{DBI}
\end{eqnarray}
\begin{widetext}
\begin{eqnarray}
            &=&\frac{f_{\pi}^2}{4}%
               \int d^4 x \mbox{tr}\lk L_{\mu}L_{\mu}\rk%
                + m_{\rho}^2%
               \int d^4 x \mbox{tr}\lk\rho_{\mu}\rho_{\mu}\rk%
                 - \frac{1}{32 e^2}%
               \int d^4 x \mbox{tr}\ldk L_{\mu}, L_{\nu}\rdk^2%
               + \frac{1}{2}%
               \int d^4 x \mbox{tr}\lk
               \partial_{\mu}\rho_{\nu}-\partial_{\nu}\rho_{\mu}\rk^2 %
               \nonumber\\
            &&+ig_{3\rho}%
               \int d^4 x \mbox{tr}\ltk
               \lk\partial_{\mu}\rho_{\nu}-\partial_{\nu}\rho_{\mu}\rk%
               \ldk \rho_{\mu}, \rho_{\nu}\rdk%
               \rtk%
                - \frac{1}{2}g_{4\rho}%
               \int d^4 x \mbox{tr}\ldk
               \rho_{\mu}, \rho_{\nu}\rdk^2%
                -i  g_1%
               \int d^4 x \mbox{tr}\ltk%
               \ldk \alpha_{\mu}, \alpha_{\nu}\rdk%
               \lk\partial_{\mu}\rho_{\nu}-\partial_{\nu}\rho_{\mu}\rk%
               \rtk%
               \nonumber\\
            &&+ g_2%
               \int d^4 x \mbox{tr}\ltk%
               \ldk \alpha_{\mu}, \alpha_{\nu}\rdk%
               \ldk \rho_{\mu}, \rho_{\nu}\rdk%
               \rtk%
               + g_3%
               \int d^4 x \mbox{tr}\ltk%
               \ldk \alpha_{\mu}, \alpha_{\nu}\rdk%
               \lk%
               \ldk \beta_{\mu}, \rho_{\nu}\rdk +%
               \ldk \rho_{\mu}, \beta_{\nu}\rdk%
               \rk%
               \rtk%
               \nonumber\\
            &&+i g_4%
               \int d^4 x \mbox{tr}\ltk
               \lk\partial_{\mu}\rho_{\nu}-\partial_{\nu}\rho_{\mu}\rk%
               \lk%
               \ldk \beta_{\mu}, \rho_{\nu}\rdk +%
               \ldk \rho_{\mu}, \beta_{\nu}\rdk%
               \rk%
               \rtk%
               - g_5%
               \int d^4 x \mbox{tr}\ltk%
               \ldk \rho_{\mu}, \rho_{\nu}\rdk%
               \lk%
               \ldk \beta_{\mu}, \rho_{\nu}\rdk +%
               \ldk \rho_{\mu}, \beta_{\nu}\rdk%
               \rk%
               \rtk%
               \nonumber\\
            &&-\frac{1}{2} g_6%
               \int d^4 x \mbox{tr}\lk
               \ldk \alpha_{\mu}, \rho_{\nu}\rdk +%
               \ldk \rho_{\mu}, \alpha_{\nu}\rdk%
               \rk^2%
            -\frac{1}{2} g_7%
               \int d^4 x \mbox{tr}\lk
               \ldk \beta_{\mu}, \rho_{\nu}\rdk +%
               \ldk \rho_{\mu}, \beta_{\nu}\rdk%
               \rk^2%
               ,\label{f11_1}
\end{eqnarray}
\end{widetext}
where $L_\mu$ is the 1-form of pion fields 
as
\begin{eqnarray}
L_\mu(x_\nu)&\equiv& \frac{1}{i}U^\dagger(x_\nu) \partial_\mu U(x_\nu). \label{U_def}
\end{eqnarray}
There exist twelve kinds of coupling constants:
$f_\pi$, $m_\rho$, $e$, $g_{3\rho}$, $g_{4\rho}$ and $g_{1\sim 7}$
in the action (\ref{f11_1}).
However all the coupling constants are uniquely determined
from the properties of meson wave functions in the extra fifth
dimension $z$, i.e.,
the complete orthogonal basis $\psi_\pm(z)$ and $\psi_1(z)$
with oscillating eigen value $\lambda_1$ as follows:

\begin{eqnarray}
\frac{f_{\pi}^2}{4}&\equiv&\kappa\int dz 
                           K(z) (\partial_z\hat{\psi}_0)^2
=\frac{\kappa}{\pi},\label{f_pi_1}\\
m_\rho^2 &\equiv& m_1^2=\lambda_1,\label{rho_mass}\\
\frac{1}{16 e^2} &\equiv& \kappa\int dz K(z)^{-1/3}\psi_+^2\lk 1-\psi_+\rk^2,\label{Sky_para}\\
g_{3\rho}   &\equiv& \kappa\int dz K(z)^{-1/3}\psi_1^3, \label{g_3rho}\\
g_{4\rho}    &\equiv& \kappa\int dz K(z)^{-1/3}\psi_1^4, \label{g_4rho}\\
g_1    &\equiv& \kappa\int dz K(z)^{-1/3}\psi_1\psi_+\psi_-, \label{g_1}\\
g_2    &\equiv& \kappa\int dz K(z)^{-1/3}\psi_1^2\lk \frac{1}{4}-\hat{\psi}_0^2\rk, \label{g_2}\\
g_3    &\equiv& \kappa\int dz K(z)^{-1/3}\psi_1\psi_+\psi_- = g_1, \label{g_3}\\
g_4    &\equiv& \kappa\int dz K(z)^{-1/3}\psi_1^2 = 1, \label{g_4}\\
g_5    &\equiv& \kappa\int dz K(z)^{-1/3}\psi_1^3 = g_{3\rho}, \label{g_5}\\
g_6    &\equiv& \kappa\int dz K(z)^{-1/3}\psi_1^2\hat{\psi}_0^2=\frac{1}{4}-g_2, \label{g_6}\\
g_7    &\equiv& \kappa\int dz K(z)^{-1/3}\psi_1^2 = 1. \label{g_7}
\end{eqnarray}
The holographic model has two parameters 
$\kappa(=\frac{\lambda N_c}{216\pi^3})$
and the Kaluza-Klein mass $M_{\rm KK}$ as the ultraviolet cutoff scale of
this theory.
$\kappa$ appears in front of the effective action (\ref{5dimDBI_2})
because the effective action of ${\rm D}8$ brane with ${\rm D}4$ supergravity
background expanded up to $O(F^2)$ corresponds to the leading order of 
$1/N_c$ and $1/\lambda$ expansions.
Therefore, by fixing two parameters
$\kappa$ and $M_{ {\rm KK} }$ to adjust 
experimental inputs for $f_\pi$ and $m_\rho$, then
all the coupling constants (\ref{f_pi_1})$\sim$(\ref{g_7}) are uniquely
determined through the background of extra fifth dimension $z$.
Such uniqueness of the action is one of the remarkable consequences
in holographic QCD.

\section{Brane-induced Skyrmion on  ${\boldmath \mbox{${\rm R}$}}^3$ and
 $S^3$ \label{BISrs}}
In this section,
we discuss baryons 
and baryonic matter
in holographic QCD.
In Sec.~\ref{BIS_R3},
we describe the baryon 
as a chiral soliton
in the four-dimensional meson effective action
$S_{\rm eff}$ in (\ref{f11_1})
including pion and $\rho$ meson fields
derived from holographic QCD.
We call this topological soliton as the ``brane-induced Skyrmion''.
The hedgehog mass of the 
brane-induced Skyrmion
on the flat coordinate space ${\boldmath \mbox{${\rm R}$}}^3$
is derived,
which is originally given in our previous 
paper~\cite{NSK}.
In Sec.~\ref{BIS_S3},
we newly discuss the baryonic matter
in holographic QCD
by analyzing the system of 
single 
brane-induced Skyrmion
on the three-dimensional closed manifold $S^3$.
Through the projection  
procedure from the flat space
${\boldmath \mbox{${\rm R}$}}^3$ onto the curved space $S^3$,
the hedgehog mass and the 
Euler-Lagrange equations of the
brane-induced Skyrmion
on $S^3$ are derived.
All the numerical results and their physical interpretations
are presented 
in Secs.~\ref{NR},~\ref{CRDPT},~\ref{CDT}.

\subsection{Brane-induced Skyrmion on ${\boldmath \mbox{${\rm R}$}}^3$ \label{BIS_R3}}
In this work, we describe the baryon 
as the four-dimensional chiral soliton, i.e., the Skyrmion in holographic QCD.
To see the validity of this approach for the baryon, 
we now 
compare the meson effective action induced by
holographic QCD with that in a chiral perturbation
theory 
(ChPT) as a low-energy effective theory of QCD~\cite{Wein}.
The 
ChPT
is phenomenologically constructed 
respecting the symmetry of QCD, 
the chiral symmetry and the Lorentz
invariance 
in the four-dimensional space-time.
With these symmetry constraints, there 
are
three possible terms as
the four-derivative terms of pion fields: 
\begin{equation}
{\rm tr}[L_\mu, L_\nu]^2,\hspace{4mm}{\rm tr}\{L_\mu, L_\nu\}^2,\hspace{4mm}
{\rm tr}(\partial_\mu  L_\nu)^2, 
\label{deri1}
\end{equation}
where $L_\mu=\frac{1}{i}U^\dagger\partial_\mu U$ 
is the 1-form pion fields in Eq.~(\ref{U_def}).
The first term ${\rm tr}[L_\mu, L_\nu]^2$, 
called the ``Skyrme term''~\cite{Sky_mode},
is to give the stability of the Skyrme soliton solution with finite size
in the coordinate space.
On the other hand, 
the other two terms are
known to give the instability of Skyrme solitons~\cite{Zahed}.
The symmetry constraints in the ChPT can not
determine which terms should appear
because all the terms in (\ref{deri1}) are chiral symmetric 
and Lorentz invariant.
Therefore Skyrme deliberately takes only 
the first term 
${\rm tr}[L_\mu, L_\nu]^2$ 
in the meson effective action
as an effective ``model'' for the baryon as the chiral soliton, 
which was called the ``Skyrme model''~\cite{Sky_mode}.

Now, by starting from the holographic QCD with the five-dimensional
Yang-Mills action (\ref{5dimDBI_2}) of the probe ${\rm D}8$ brane,
one can
find only the Skyrme term
without the other two in (\ref{deri1})
, which is manifestly seen in the action (\ref{f11_1}).
Actually the five-dimensional
Yang-Mills action (\ref{5dimDBI_2}) with $O(F^2)$
includes two-time derivatives at most,
so that the appearances of the other two terms: 
${\rm tr}\{L_\mu, L_\nu\}^2$ and ${\rm tr}(\partial_\mu  L_\nu)^2$
with four-time derivatives are 
forbidden 
at the leading order of $1/N_c$ 
and $1/\lambda$ expansion in holographic QCD.
These comparisons with the chiral perturbation theory
clearly indicate that
holographic QCD is not just the low-energy effective theory
of QCD only with the constraint of symmetries in four-dimensional
space-time: actually, it obeys 
the $U(N_f)$ symmetry extending to the extra fifth dimension $z$.
%
Furthermore, 
one can see that
the chiral soliton picture for the baryon is now supported by 
the holographic approach, retaining the direct connection with QCD.
With these considerations,
we employ the concept of chiral soliton picture
for the baryon analysis in holographic QCD~\cite{NSK}.

Now we begin with the hedgehog Ansatz
for pion field $U({\bf x})$ 
and $\rho$ meson field $\rho_{\mu}({\bf x})$
as a baryon configuration~\cite{NSK}:
\begin{eqnarray}
&&U^{\star}({\bf x})=e^{i\tau_a \hat{x}_a F(r)}, 
\hspace{5mm}\mbox{($\hat{x}_a\equiv \frac{x_a}{r}$, $r\equiv|{\bf x}|$)}\label{HH}\\
&&\rho^{\star}_{0}({\bf x})=0, \hspace{5mm} \rho^{\star}_{i}({\bf x})=\rho^{\star}_{i a}({\bf x})\frac{\tau_a}{2}
                                               =\ltk \varepsilon_{i a
                                               b}\hat{x}_b\tilde{G}(r)\rtk
                                               \tau_a,\nonumber\\
&&                                      \hspace{45mm}(\tilde{G}(r)\equiv
                                               G(r)/r)\label{WYTP}
\end{eqnarray}
where $\tau_a$ is Pauli matrix, and 
$F(r)$ is a dimensionless profile function of the pion field
with boundary conditions $F(0)=\pi$ and $F(\infty)=0$, giving
topological charge equal to unity. 
Ansatz (\ref{HH}) means $\pi_a({\bf x})=\hat{x}_a F(r)$ for the pion
field.
$G(r)$ is also a dimensionless profile function of the $\rho$ meson field.
This Ansatz for $\rho$ meson field is also 
called ``Wu-Yang-'tHooft-Polyakov
Ansatz''~\cite{Igarashi}, and
the same configuration Ansatz can be seen for the gauge field 
of the 'tHooft-Polyakov monopole~\cite{Raja}.

By substituting the configuration Ansatz 
(\ref{HH}) and
(\ref{WYTP}) into the four-dimensional meson effective action
$S_{\rm eff}$ in (\ref{f11_1}) with the Euclidean metric,
we can  
get the static hedgehog mass of a brane-induced Skyrmion 
on the flat space
${\boldmath \mbox{${\rm R}$}}^3$ as follows 
(detailed derivations 
 can be found in our previous paper~\cite{NSK}):
\begin{widetext}
\begin{eqnarray}
E[F(r), G(r)]&\equiv& 
             \ldk S_{{\rm D}8}^{\rm DBI}-S_{{\rm D}8}^{\rm DBI}|_{A_M\rightarrow 0}
             \rdk_{{\rm hedgehog}}
                     \equiv \int_0^{\infty}4\pi dr r^2\cdot\varepsilon [F(r),
                     G(r)], \label{BIS_energy}\\
r^2\cdot\varepsilon [F(r), G(r)]%
          &=&%
             \frac{f_\pi^2}{4}\ldk 2\lk r^2 F'^{2}+2\sin^2F \rk\rdk
%
           + m_{\rho}^2%
             \ldk 4 r^2 \tilde{G}^2\rdk
%
           + \frac{1}{32e^2}\ldk 16 \sin^2F \lk 2 F'^{2}+\frac{\sin^2F}{r^2}\rk\rdk 
               \nonumber \\ 
          &+&\frac{1}{2}%
             \ldk 8\ltk
             3\tilde{G}^2+2 r \tilde{G}(\tilde{G}') +r^2\tilde{G}'^{2}\rtk\rdk
%
           -  g_{3\rho}  %
             \ldk 16 r \tilde{G}^3 \rdk
%
           +  \frac{1}{2}g_{4\rho}%
             \ldk 16 r^2 \tilde{G}^4 \rdk
               \nonumber \\
          &+&  g_1 %
             \ldk 16 \ltk F'\sin F \cdot\lk
             \tilde{G}+ r \tilde{G}'\rk%
             +\sin^2F\cdot\tilde{G}/r\rtk\rdk
%
           -  g_2%
             \ldk 16 \sin^2F\cdot\tilde{G}^2 \rdk
               \nonumber \\
          &-& g_3 %
             \ldk 16 \sin^2F\cdot\lk 1-\cos F\rk \tilde{G}/r \rdk 
%
           -  g_4%
             \ldk 16 \lk 1-\cos F\rk\tilde{G}^2\rdk   
%
           +  g_5 %
             \ldk 16 r \lk 1-\cos F \rk \tilde{G}^3\rdk
               \nonumber  \\
          &+& g_6%
             \ldk 16 r^2 F'^{2}\tilde{G}^2\rdk
%
           +  g_7%
             \ldk 8 \lk 1-\cos F\rk^2\tilde{G}^2\rdk,\label{energy_dense_An}
\end{eqnarray}
where $F'\equiv\frac{dF(r)}{dr}(=\frac{\partial F(r)}{\partial r})$ 
and      
$\tilde{G}'\equiv\frac{d\tilde{G}(r)}{dr}(=\frac{\partial \tilde{G}(r)}{\partial r})$. 

Now we take the ``Adkins-Nappi-Witten (ANW) unit'' for energy and length
as $E_{\rm ANW}\equiv \frac{f_\pi}{2e}$ and 
$r_{\rm ANW}\equiv \frac{1}{ef_\pi}$~\cite{ANW},
and we rewrite all variables in this ANW unit as
$\overline{E}\equiv \frac{1}{E_{\rm ANW}}E$
and
$\overline{r}\equiv \frac{1}{r_{\rm ANW}}r$.
By taking this scaled unit,
the hedgehog energy density (\ref{energy_dense_An}) of single
brane-induced Skyrmion on ${\boldmath \mbox{${\rm R}$} }^3$
can be rewritten
as follows (overlines of $\overline{E}$ and $\overline{r}$ below are
abbreviated for simplicity): 
\begin{eqnarray}
r^2\cdot\varepsilon [F(r), G(r)]%
          &=&%
             \lk r^2 F'^{2}+2\sin^2F \rk
%
           + 2 \lk\frac{m_{\rho}}{f_{\pi}}\rk^2%
             \ldk 4 r^2 \tilde{G}^2\rdk
%
           + \sin^2F \lk 2 F'^{2}+\frac{\sin^2F}{r^2}\rk 
               \nonumber 
\end{eqnarray}
\begin{eqnarray} 
          &+&\lk 2e^2\rk \frac{1}{2}%
             \ldk 8\ltk
             3\tilde{G}^2+2 r \tilde{G}(\tilde{G}') +r^2\tilde{G}'^{2}\rtk\rdk
%
           -  \lk 2e^2\rk g_{3\rho}  %
             \ldk 16 r \tilde{G}^3 \rdk
%
           + \lk 2e^2\rk \frac{1}{2}g_{4\rho}%
             \ldk 16 r^2 \tilde{G}^4 \rdk
               \nonumber \\
          &+&\lk 2e^2\rk  g_1 %
             \ldk 16 \ltk F'\sin F \cdot\lk
             \tilde{G}+ r \tilde{G}'\rk%
             +\sin^2F\cdot\tilde{G}/r\rtk\rdk
%
           - \lk 2e^2\rk g_2%
             \ldk 16 \sin^2F\cdot\tilde{G}^2 \rdk
               \nonumber \\
          &-&\lk 2e^2\rk g_3 %
             \ldk 16 \sin^2F\cdot\lk 1-\cos F\rk \tilde{G}/r \rdk 
%
           - \lk 2e^2\rk g_4%
             \ldk 16 \lk 1-\cos F\rk\tilde{G}^2\rdk   
               \nonumber \\
          &+& \lk 2e^2\rk g_5 %
             \ldk 16 r \lk 1-\cos F \rk \tilde{G}^3\rdk
%
           + \lk 2e^2\rk g_6%
             \ldk 16 r^2 F'^{2}\tilde{G}^2\rdk
%
           + \lk 2e^2\rk g_7%
             \ldk 8 \lk 1-\cos F\rk^2\tilde{G}^2\rdk.\label{energy_dense_Re_R3}
\end{eqnarray}         
\end{widetext}

Here we comment about a scaling property of the 
brane-induced Skyrmion.
The holographic QCD has just two parameters:
$\kappa(=\frac{\lambda N_c}{216\pi^3})$ and $M_{\rm KK}$, so that 
the pion decay constant $f_\pi$, the $\rho$ meson mass $m_\rho$, and
the Skyrme parameter $e$ in Eqs.(\ref{f_pi_1})$\sim$(\ref{Sky_para})
can be explicitly written by $\kappa$ and $M_{\rm KK}$
in holographic QCD as 
\begin{eqnarray}
f_{\pi}&=&2\sqrt{\frac{\kappa}{\pi}}M_{\rm KK},\label{fpi_Mkk}\\
m_\rho&=&\sqrt{\lambda_1}M_{\rm KK}\simeq \sqrt{0.67}M_{\rm
 KK},\label{mrho_Mkk}\\
e&=&\frac{1}{4}\ldk \kappa\int dz K^{-1/3}
                  \psi_{+}^2(1-\psi_{+})^2 \rdk ^{-1/2}
  \simeq \frac{1}{4\sqrt{0.157}}\frac{1}{\sqrt{\kappa}},\nonumber\\
\label{e_kappa}
\end{eqnarray}
where the energy unit $M_{\rm KK}$ is recovered.
By using these relations (\ref{fpi_Mkk})$\sim$(\ref{e_kappa}),
ANW unit for energy and length, i.e., $E_{\rm ANW}$ and $r_{\rm ANW}$
can be written by $\kappa$ and $M_{\rm KK}$ as 
\begin{eqnarray}
E_{\rm ANW}=\frac{f_\pi}{2e}={\rm const}\cdot \kappa M_{\rm KK},\label{ANW_holo1}\\
r_{\rm ANW }=\frac{1}{ef_\pi}={\rm const}\cdot\frac{1}{M_{\rm KK}}.\label{ANW_holo2}
\end{eqnarray}
The five-dimensional Yang-Mills action (\ref{5dimDBI_2}) 
with $O(F^2)$
is proportional to $\kappa(=\frac{\lambda N_c}{216\pi^3})$, 
as the leading order of $1/N_c$ and $1/\lambda$ expansion.
Furthermore $M_{\rm KK}$ is the sole energy scale of the holographic
approach.
Therefore, in the energy unit $E_{\rm ANW}(\propto \kappa M_{\rm KK})$,
the total energy appears as a scale invariant variable. 
In fact,
by introducing the rescaled $\rho$ meson field $\widehat{G}(r)$ as
\begin{eqnarray}
\widehat{G}(r)\equiv \frac{1}{\sqrt{\kappa}}\tilde{G}(r),\label{widehatG}
\end{eqnarray}
and considering the $\kappa$-dependence of the basis $\psi_1$
as $\psi_1\propto \frac{1}{\sqrt{\kappa}}$
in the normalization condition (\ref{normal_psi_n}),
one can analytically show that 
every energy density in each term of Eq.(\ref{energy_dense_Re_R3})
and meson field configurations $F(r)$ and $\widehat{G}(r)$ are
scale invariant variables,
being independent of the holographic two parameters, $\kappa$ and 
$M_{\rm KK}$.

With the considerations above,
we give most discussions below
in the ANW unit
as the universal features of 
baryonic matter in holographic QCD,
being independent of the definite values
of $f_\pi$ and $m_\rho$.
The recovering of physical unit with the experimental inputs
for $f_\pi$ and $m_\rho$ is discussed in Sec.~\ref{CDT},
with respect to the critical densities of the phase transitions
in the baryonic matter within holographic approach.

\subsection{Brane-induced Skyrmion on $S^3$ \label{BIS_S3}}
Now we study the baryonic matter in holographic QCD
by analyzing the system of single brane-induce Skyrmion
on a three-dimensional closed manifold $S^3$.

In this study, we consider the baryonic matter with large $N_c$
because holographic QCD is 
the large $N_c$ effective theory, derived from the classical supergravity 
	justified in the large $N_c$ and large 'tHooft coupling~\cite{Kru}.
According to the general analysis of large-$N_c$ QCD,
a baryon mass is found to become $O(N_c)$~\cite{tH, Witten},
so that its kinetic energy becomes $O(N_c^{-1})$.
As for the quantum effects of the baryonic matter, 
zero point quantum fluctuation energy $E_0$ and 
baryon mass splitting $\Delta m$ in the isospin projection
correspond to the higher-order effects of
$1/N_c$ expansion:
$E_0\sim O(N_c^0)$ and $\Delta m\sim O(N_c^{-1})$~\cite{ANW}.
Therefore, with large-$N_c$ condition,
we can consider that
the kinetic energy and quantum effects are suppressed 
relative to the static mass, and 
the baryonic matter comes into the ``static Skyrme matter''. 

Such static Skyrme matter was first analyzed by Klebanov~\cite{Kle},
placing Skyrme soliton configurations periodically
along the three-dimensional cubic lattice,
which could be related with ``nuclear crystal'' with pion condensation
in the deep interior of neutron stars.
Therefore, by analyzing the static Skyrme matter,
one can see some typical features of baryonic matter
with large-$N_c$ conditions.

\begin{figure}[h]
  \begin{center}
       \resizebox{83mm}{!}{\includegraphics{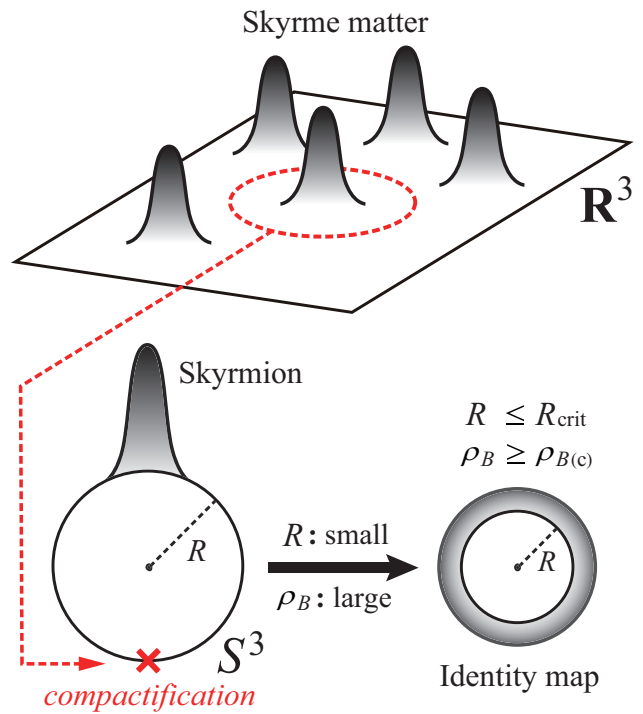}}\\
  \end{center}
\caption{Schematic figure of the static Skyrme matter on a flat coordinate space ${\boldmath \mbox{${\rm R}$}}^3$,
and the system of single Skyrmion on a closed manifold $S^3$ with finite radius $R$.
The static Skyrme matter on ${\boldmath \mbox{${\rm R}$}}^3$
and 
the system of single Skyrmion on $S^3$ can be related with each other
through the compactification of the boundary for a unit cell on
${\boldmath \mbox{${\rm R}$}}^3$
shared by the single Skyrmion.
The decrease of 
the radius $R$ of $S^3$ represents 
the increase of the baryon-number density $\rho_B(\equiv (2\pi^2 R^3)^{-1})$ 
in the medium in this modeling.
For $R\leq R_{\rm crit}$ as a critical radius, i.e.,
$\rho_{B}\geq\rho_{B({\rm c})}(\equiv \{2\pi^2 (R_{\rm crit})^3\}^{-1})$ 
as a critical density, 
the energy density of the single Skyrmion becomes
uniform distribution as the 
``identity map'' discussed in Eq.~(\ref{BIS_ide}),
which is called the ``delocalization phase transition''.}
  \label{fig_projection1}
\end{figure}

In this paper, 
we take certain
mathematical trick to analyze
such static Skyrme matter suggested by Manton and Ruback~\cite{MR}.
To analyze the features of multi-Skyrmion system
on the flat coordinate space 
${\boldmath \mbox{${\rm R}$}}^3$,
they alternately treat the system of 
single Skyrmion
on a three-dimensional closed manifold $S^3$
with finite radius $R$ as shown in Fig.~\ref{fig_projection1}.
Actually,
the multi-Skyrmion system on
${\boldmath \mbox{${\rm R}$}}^3$
and the system of single Skyrmion
on $S^3$
can be related with each other 
through the compactification of the boundary for a unit cell on
${\boldmath \mbox{${\rm R}$}}^3$
shared by the single Skyrmion as in Fig.~\ref{fig_projection1}.
The interaction between the baryons in the medium on
${\boldmath \mbox{${\rm R}$}}^3$
is simulated by the curvature of the closed manifold $S^3$.
The baryon-number density can be 
represented 
as $\rho_B=1/2\pi^2 R^3$ on $S^3$,
so that, as the size of $S^3$ decreases,
the increase of
the baryon-number density 
in the medium is represented
in this modeling.
Actually,
as the radius $R$ of $S^3$ decreases,
the energy density of single baryon is found to delocalize 
due to the 
medium effects 
in the baryonic matter~\cite{MR},
and, below the critical radius $R_{\rm crit}$ of $S^3$,
the energy density of the baryon coincides with the
uniform distribution as the ``identity map'',
which is called the ``delocalization phase transition'' 
shown in Fig.~\ref{fig_projection1}.
Such delocalization phase transition
in the Skyrme model can
be related with the deconfinement of the baryon
and also the chiral symmetry restoration in QCD, which 
will be
inclusively discussed by taking the order parameters 
in Sec.~\ref{CRDPT}.
With these considerations,
by analyzing the system of single 
brane-induced Skyrmion on 
$S^3$, we can see some typical features of baryonic matter
in holographic QCD.
Especially, by comparing the standard Skyrmion 
without $\rho$ mesons and the brane-induced Skyrmion on 
$S^3$,
the roles of (axial) vector mesons in high density phase of baryonic matter
can be discussed from the holographic point of view.

Now we introduce the projection procedure from
the flat space 
${\boldmath \mbox{${\rm R}$}}^3$
onto the curved space $S^3$~\cite{SaSu},
to get a hedgehog mass of a brane-induced Skyrmion on $S^3$.
First, we consider 
the three-dimensional orthogonal space
${\boldmath \mbox{${\rm R}$}}^3$
in polar coordinates as 
\begin{eqnarray}
{\bf x}&=&(z, x, y)\nonumber \\
       &=&(r\cos\theta, r\sin\theta\cos\phi,
       r\sin\theta\sin\phi) \nonumber\\
       &=&(r, \theta, \phi)_{\rm 3dim.polar}.\label{coordR3}
\end{eqnarray}
The integral operator $d\hat{\bf x}$ and the derivation $d$
can be written in polar coordinates as
\begin{eqnarray}
d\hat{\bf x}&=&(d\hat{r}, d\hat{\theta}, d\hat{\phi})_{\rm 3dim.polar}
\nonumber\\
&=&
(dr, rd\theta, r\sin\theta d\phi)_{\rm 3dim.polar},\label{3dimIO}\\
d &=&dr\frac{\partial}{\partial r}+
  d\theta\frac{\partial}{\partial\theta}+
  d\phi\frac{\partial}{\partial\phi}\nonumber\\
  &=&d\hat{r}\frac{\partial}{\partial r}+
     d\hat{\theta}\frac{1}{r}\frac{\partial}{\partial\theta}+
     d\hat{\phi}\frac{1}{r\sin\theta}\frac{\partial}{\partial\phi},
     \label{3dimDERI}
\end{eqnarray}
so that the differential operator ${\boldmath \mbox{$\partial$}}$
can be written as
\begin{eqnarray}
{\boldmath \mbox{$\partial$}}=\lk\frac{\partial}{\partial r},
                  \frac{1}{r}\frac{\partial}{\partial\theta},
                  \frac{1}{r\sin\theta}\frac{\partial}{\partial\phi}                           
                  \rk_{\rm 3dim.polar}.\label{3dimDO}
\end{eqnarray}

Second, we consider 
the four-dimensional orthogonal space
${\boldmath \mbox{${\rm R}$}}^4$
in polar coordinates as 
\begin{eqnarray}
{\bf X}&=&(t, z, x, y)\nonumber \\
       &=&(R\cos\Theta, R\sin\Theta\cos\theta, R\sin\Theta\sin\theta\cos\phi,
                  \nonumber\\
        &&   R\sin\Theta\sin\theta\sin\phi) \nonumber\\
       &=&(R, \Theta, \theta, \phi)_{\rm 4dim.polar}.\label{coordR4}
\end{eqnarray}
The integral operator $d\hat{\bf x}$ and the derivation $d$
can be written in 
polar coordinates as  
\begin{eqnarray}
d\hat{\bf x}&=&(d\hat{R}, d\hat{\Theta}, d\hat{\theta}, d\hat{\phi})_{\rm 4dim.polar}
\nonumber\\
&=&(dR, Rd\Theta, R\sin\Theta d\theta, R\sin\Theta\sin\theta d\phi)_{\rm 4dim.polar},
\nonumber\\
\label{4dimIO}\\
d &=&dR\frac{\partial}{\partial R}+
  d\Theta\frac{\partial}{\partial\Theta}+
  d\theta\frac{\partial}{\partial\theta}+
  d\phi\frac{\partial}{\partial\phi}\nonumber\\
  &=&d\hat{R}\frac{\partial}{\partial R}+
     d\hat{\Theta}\frac{1}{R}\frac{\partial}{\partial\Theta}+
     d\hat{\theta}\frac{1}{R\sin\Theta}\frac{\partial}{\partial\theta}
    \nonumber\\
 &&+d\hat{\phi}\frac{1}{R\sin\Theta\sin\theta}\frac{\partial}{\partial\phi},
     \label{4dimDERI}
\end{eqnarray}
so that the differential operator ${\boldmath \mbox{$\partial$}}$
can be written as
\begin{eqnarray}
{\boldmath \mbox{$\partial$}}=\lk
               \frac{\partial}{\partial R},
               \frac{1}{R}\frac{\partial}{\partial\Theta},
               \frac{1}{R\sin\Theta}\frac{\partial}{\partial\theta},
               \frac{1}{R\sin\Theta\sin\theta}\frac{\partial}{\partial\phi}
               \rk_{\rm 4dim.polar}.\nonumber\\
              \hspace*{-5mm}\label{4dimDO}
\end{eqnarray}

Now, by limiting the four-dimensional orthogonal space 
${\boldmath \mbox{${\rm R}$}}^4$ 
onto the surface of a three-dimensional closed manifold $S^3$ with fixed
radius $R$,
the coordinate $t$ in Eq.~(\ref{coordR4}) becomes dependent on the 
other coordinates $(z, x, y)$.
Furthermore, $dR$ and
$\frac{\partial}{\partial R}$
can be regarded as zero in Eqs.~(\ref{4dimIO}) and (\ref{4dimDO})
because the radial coordinate $R$ is fixed on $S^3$.
Therefore, by comparing Eqs.~(\ref{coordR3}), (\ref{3dimIO}),
(\ref{3dimDO}) on ${\boldmath \mbox{${\rm R}$}}^3$ , and
Eqs.~(\ref{coordR4}), (\ref{4dimIO}), (\ref{4dimDO}) on $S^3$ with fixed
radius $R$,
we find the projection procedure from  
${\boldmath \mbox{${\rm R}$}}^3$ to $S^3$ as follows:
\begin{eqnarray}
r&\longrightarrow&R\sin\Theta,\label{pro1}\\
dr&\longrightarrow&Rd\Theta,\label{pro2}\\
\frac{\partial}{\partial r}&\longrightarrow&
              \frac{1}{R}\frac{\partial}{\partial \Theta}.\label{pro3}
\end{eqnarray}

Recall that
the hedgehog mass on 
${\boldmath \mbox{${\rm R}$}}^3$ 
with energy density in Eq.~(\ref{energy_dense_Re_R3})
can be written with its explicit arguments for the energy density as
\begin{eqnarray}
E= \int_0^{\infty}4\pi dr r^2\cdot\varepsilon \ldk F(r), G(r), 
        \frac{\partial}{\partial r}F(r), 
        \frac{\partial}{\partial r}G(r), r\rdk. \nonumber\\
        \label{BIS_energy_R3}
\end{eqnarray}
%
By applying the projection procedure (\ref{pro1}), (\ref{pro2}) and
(\ref{pro3}) to
the hedgehog mass on ${\boldmath \mbox{${\rm R}$}}^3$
in Eq.~(\ref{BIS_energy_R3}),
we can get the hedgehog mass on $S^3$ as 
\begin{widetext}
\begin{eqnarray}
E&=& \int_0^{\pi}4\pi Rd\Theta R^2\sin^2\Theta\cdot\varepsilon \ldk F(R\sin\Theta), G(R\sin\Theta), 
        \frac{1}{R}\frac{\partial}{\partial \Theta}F(R\sin\Theta), 
        \frac{1}{R}\frac{\partial}{\partial \Theta}G(R\sin\Theta), R\sin\Theta\rdk. \label{BIS_energy_S3}\\
 &=& \int_0^{\pi R}4\pi dr R^2\sin^2\frac{r}{R}\cdot\varepsilon \ldk F(r), G(r), 
        \frac{\partial}{\partial r}F(r), 
        \frac{\partial}{\partial r}G(r), R\sin\frac{r}{R}\rdk. \label{BIS_energy_S3_re}
\end{eqnarray}
In Eq.~(\ref{BIS_energy_S3_re}), we introduce a new variable $r$ 
as the arc length on $S^3$ as
\begin{eqnarray}
r\equiv R\Theta, 
\end{eqnarray}
and $r$-dependent dimensionless functions 
$F(R\sin\frac{r}{R})$ and $G(R\sin\frac{r}{R})$ are renamed again as
$F(r)$ and $G(r)$.
Therefore, by comparing Eq.~(\ref{BIS_energy_R3}) on
${\boldmath \mbox{${\rm R}$}}^3$
and Eq.~(\ref{BIS_energy_S3_re}) on $S^3$,
we can get the simple projection procedure for the hedgehog energy density from
${\boldmath \mbox{${\rm R}$}}^3$ to $S^3$ as 
\begin{eqnarray}
dr r^2\cdot\varepsilon \ldk F(r), G(r), 
        \frac{\partial}{\partial r}F(r), 
        \frac{\partial}{\partial r}G(r), r\rdk
\longrightarrow 
dr R^2\sin^2\frac{r}{R}\cdot\varepsilon \ldk F(r), G(r), 
        \frac{\partial}{\partial r}F(r), 
        \frac{\partial}{\partial r}G(r), R\sin\frac{r}{R}\rdk, \label{Edensity_pro1}
\end{eqnarray}
where the topological boundary for the chiral field $F(r)$ in (\ref{HH})
is also projected on $S^3$ as
\begin{eqnarray}
F(0)=\pi, \hspace{3mm}F(\pi R)=0. \label{bondary_S3}
\end{eqnarray}

Now, by applying
the projection procedure (\ref{Edensity_pro1})
for the hedgehog energy density 
on ${\boldmath \mbox{${\rm R}$}}^3$ in Eq.~(\ref{energy_dense_Re_R3}),
we can eventually get
the hedgehog energy density on $S^3$ 
with ANW units as follows
(Note here that the dimensional profile function $\tilde{G}(r)=G(r)/r$
on ${\boldmath \mbox{${\rm R}$}}^3$
is to be naturally introduced on $S^3$ through the projection procedure as 
$\tilde{G}(r)\equiv \frac{G(r)}{R\sin\frac{r}{R}}$):
\begin{eqnarray}
E[F(r), G(r)]&=& 
                     \int_0^{\pi R}4\pi dr R^2\sin^2\frac{r}{R}\cdot\varepsilon [F(r),
                     G(r)], \label{BIS_energy_S3_again}\\
R^2\sin^2\frac{r}{R}\cdot\varepsilon [F(r), G(r)]%
          &=&%
             \lk R^2\sin^2\frac{r}{R}\cdot F'^{2}+2\sin^2F \rk
%
           + 2 \lk\frac{m_{\rho}}{f_{\pi}}\rk^2%
             \ldk 4 R^2\sin^2\frac{r}{R}\cdot \tilde{G}^2\rdk
               \nonumber  \\ 
          &+&%
               \sin^2F \lk 2 F'^{2}+\frac{\sin^2F}{R^2\sin^2\frac{r}{R}}\rk 
               \nonumber \\ 
          &+&\lk 2e^2\rk \frac{1}{2}%
             \ldk 8\ltk
             \lk 2+\cos^2\frac{r}{R}\rk\tilde{G}^2+
             2R\sin\frac{r}{R}\cos\frac{r}{R}\cdot\tilde{G}(\tilde{G}')
             +R^2\sin^2\frac{r}{R}\cdot\tilde{G}'^{2}\rtk\rdk
               \nonumber \\
          &-& \lk 2e^2\rk g_{3\rho}  %
             \ldk 16 R\sin\frac{r}{R}\cdot \tilde{G}^3 \rdk
%
           + \lk 2e^2\rk \frac{1}{2}g_{4\rho}%
             \ldk 16 R^2\sin^2\frac{r}{R}\cdot  \tilde{G}^4 \rdk
               \nonumber \\
          &+&\lk 2e^2\rk  g_1 %
             \ldk 16 \ltk F'\sin F \cdot\lk
             \cos\frac{r}{R}\cdot\tilde{G}+ R\sin\frac{r}{R}\cdot \tilde{G}'\rk%
             +\sin^2F\cdot\tilde{G}/\lk R\sin\frac{r}{R}\rk\rtk\rdk
               \nonumber \\
          &-&\lk 2e^2\rk g_2%
             \ldk 16 \sin^2F\cdot\tilde{G}^2 \rdk
%
           - \lk 2e^2\rk g_3 %
             \ldk 16 \sin^2F\cdot\lk 1-\cos F\rk \tilde{G}/\lk R\sin\frac{r}{R}\rk \rdk 
               \nonumber \\
          &-&\lk 2e^2\rk g_4%
             \ldk 16 \lk 1-\cos F\rk\tilde{G}^2\rdk   
%
           + \lk 2e^2\rk g_5 %
             \ldk 16 R\sin\frac{r}{R}\cdot \lk 1-\cos F \rk \tilde{G}^3\rdk
               \nonumber  \\
          &+&\lk 2e^2\rk g_6%
             \ldk 16 R^2\sin^2\frac{r}{R}\cdot F'^{2}\tilde{G}^2\rdk
%
           + \lk 2e^2\rk g_7%
             \ldk 8 \lk 1-\cos F\rk^2\tilde{G}^2\rdk,\label{energy_dense_Re}
\end{eqnarray}
where $F'\equiv\frac{dF(r)}{dr}(=\frac{\partial F(r)}{\partial r})$ 
and      
$\tilde{G}'\equiv\frac{d\tilde{G}(r)}{dr}(=\frac{\partial \tilde{G}(r)}{\partial r})$. 
Here 
measures 
$R^2\sin^2\frac{r}{R}$ newly appear 
in the energy density 
(\ref{energy_dense_Re})
relative to Eq.(\ref{energy_dense_Re_R3}) on the flat space
${\boldmath \mbox{${\rm R}$}}^3$, 
indicating the existence of the curvature of the closed manifold $S^3$.
We also construct the 
Euler-Lagrange equations 
for the pion field $F(r)$ 
and $\rho$ meson field $\tilde{G}(r)$ from the energy density 
in Eq.~(\ref{energy_dense_Re})
as follows:
\begin{eqnarray}
&&\frac{1}{4\pi}\ltk \frac{\delta E}{\delta F(r)}-%
                   \frac{d}{dr}\lk \frac{\delta E}{\delta
                   F'(r)}\rk \rtk \nonumber\\
&=&\lk -4R\sin\frac{r}{R}\cos\frac{r}{R}\cdot%
 F'-2R^2\sin^2\frac{r}{R}\cdot F''+4\sin F\cdot \cos F \rk \nonumber\\
&+&\ltk -4\sin F\cdot\cos F\cdot F'^{2}-4\sin^2F\cdot F''+%
                               4\sin^3F\cdot \cos F/\lk R^2\sin^2\frac{r}{R} \rk\rtk \nonumber\\                
&+& \lk 2e^2\rk g_1 \ldk 16 %
               \ltk 2\sin F\cdot\cos F\cdot\tilde{G}/\lk
               R\sin\frac{r}{R}\rk-%
                    \sin F\cdot \lk %
               -\frac{1}{R}\sin\frac{r}{R}\cdot\tilde{G}+ 
               2 \cos\frac{r}{R}\cdot\tilde{G}'+R\sin\frac{r}{R}\cdot\tilde{G}''\rk%
               \rtk \rdk \nonumber\\
&-& \lk 2e^2\rk g_2 \ldk 16 %
               \lk 2\sin F\cdot\cos F\cdot \tilde{G}^2\rk\rdk
%
 -  \lk 2e^2\rk g_3 \ldk 16 %
               \lk \sin F + 2\sin F \cdot \cos F-3\sin F\cdot \cos^2F\rk
               \tilde{G}/\lk R\sin\frac{r}{R}\rk\rdk
               \nonumber\\
&-& \lk 2e^2\rk g_4 \ldk 16 %
               \lk \sin F \cdot\tilde{G}^2\rk \rdk
%
 +  \lk 2e^2\rk g_5 \ldk 16 %
               \lk R\sin\frac{r}{R} \sin F \cdot\tilde{G}^3\rk \rdk
               \nonumber\\
&+& \lk 2e^2\rk g_6 \ldk 16 %
                \lk -4R\sin\frac{r}{R}\cos\frac{r}{R}\cdot F'\tilde{G}^2%
                    -2R^2\sin^2\frac{r}{R}\cdot F''\tilde{G}^2%
                    -4R^2\sin^2\frac{r}{R}\cdot
		    F'\tilde{G}\tilde{G}'\rk\rdk
                    \nonumber\\
&+& \lk 2e^2\rk g_7 \ldk 8 %
               \ltk 2 \lk 1-\cos F \rk \sin F \cdot\tilde{G}^2\rtk \rdk
%
 =  0,  \label{EL_F}                 
\end{eqnarray}
\begin{eqnarray}
&&\frac{1}{4\pi}\ltk \frac{\delta E}{\delta \tilde{G}(r)}-%
                   \frac{d}{dr}\lk \frac{\delta E}{\delta
                   \tilde{G}'(r)}\rk \rtk \nonumber\\
&=& 2\lk\frac{m_{\rho}}{f_{\pi}}\rk^2 \ldk 4 \lk 2
    R^2\sin^2\frac{r}{R}\cdot \tilde{G} \rk \rdk
    \nonumber\\
&+& \lk 2e^2\rk\frac{1}{2} \ldk 8 \lk 4\tilde{G}%
            +2\sin^2\frac{r}{R}\cdot\tilde{G}%
            -4R\sin\frac{r}{R}\cos\frac{r}{R}\cdot\tilde{G}'-2R^2\sin^2\frac{r}{R}\cdot\tilde{G}''\rk
            \rdk
            \nonumber\\
&-& \lk 2e^2\rk g_{3\rho} \ldk 16 \lk 3R\sin\frac{r}{R}\cdot\tilde{G}^2
    \rk \rdk
%
 +  \lk 2e^2\rk \frac{1}{2} g_{4\rho} \ldk 16 \lk 4
    R^2\sin^2\frac{r}{R}\cdot \tilde{G}^3 \rk \rdk 
    \nonumber\\
&+& \lk 2e^2\rk g_1 \ldk 16 %
               \ltk \sin^2 F/\lk R\sin\frac{r}{R}\rk-R\sin\frac{r}{R}\cos F\cdot F'^{2}-%
               R\sin\frac{r}{R}\sin F\cdot F''\rtk \rdk
               \nonumber\\
&-& \lk 2e^2\rk g_2 \ldk 16 %
               \lk 2\sin^2F \cdot\tilde{G} \rk\rdk
%
 -  \lk 2e^2\rk g_3 \ldk 16 %
                \sin^2 F (1-\cos F)/\lk R\sin\frac{r}{R}\rk \rdk
                \nonumber\\
&-& \lk 2e^2\rk g_4 \ldk 16 %
                \ltk 2(1-\cos F)\tilde{G}\rtk \rdk
%
 +  \lk 2e^2\rk g_5 \ldk 16 %
               \ltk 3R\sin\frac{r}{R}\lk 1-\cos F\rk\tilde{G}^2 \rtk
               \rdk
               \nonumber\\
&+& \lk 2e^2\rk g_6 \ldk 16 %
                \ltk 2R^2\sin^2\frac{r}{R}\cdot F'^{2}\tilde{G}\rtk\rdk
%
 +  \lk 2e^2\rk g_7 \ldk 8 %
               \ltk 2 \lk 1-\cos F \rk^2 \tilde{G}\rtk \rdk
%
 =  0 .\label{EL_G}
\end{eqnarray}
\end{widetext}

Now one can show that the 
Euler-Lagrange equations 
(\ref{EL_F}) and (\ref{EL_G}) always
have the analytical solution
for arbitrary value of radius $R$ of $S^3$ as 
\begin{eqnarray}
F(r)=\pi-\frac{r}{R},\hspace{3mm} \tilde{G}(r)=0, \label{BIS_ide}
\end{eqnarray}
which is the ``identity map'' solution in the standard Skyrme model
without $\rho$ meson field~\cite{MR}.
In the Skyrme model,
the energy density of the baryon is generated by 
the spatial gradient of the pion fields.
Therefore, such linear configuration as
in Eq.(\ref{BIS_ide})
gives the uniform energy distributions 
with its classical mass as
\begin{eqnarray}
E_{\rm id}=\lk R+\frac{1}{R}\rk 6\pi^2, \label{E_I}
\end{eqnarray}
which can be given by substituting the linear configuration 
(\ref{BIS_ide}) into the energy density 
(\ref{energy_dense_Re}).
Hence 
the change of absolute minimum solution
from the localized Skyrmion
into the identity map (\ref{BIS_ide}) 
can be regarded as a signal of transitions
from localized phase to the uniform phase
of the baryonic matter, referred 
as the delocalization phase transition.
Note
that the identity map solution (\ref{BIS_ide}) has no $\rho$ meson
configuration: $\tilde{G}(r)=0$.
Therefore, if the identity map  
is realized as the absolute minimum
solution, it indicates that
$\rho$ meson field absolutely disappear in high density phase, 
which is discussed with numerical results in Sec.~\ref{NR}. 

\section{Baryon nature in dense QCD \label{NR}}

In this section, we show the numerical results 
and discussions
about the baryon nature in dense QCD
by solving the 
Euler-Lagrange 
equations (\ref{EL_F}) and (\ref{EL_G}) derived from holographic QCD.
The energy density 
and the field configuration profiles of the baryon
are analyzed in Sec.~\ref{ECpro}.
A new striking picture of ``pion dominance''
near the critical density is proposed in Sec.~\ref{PDNC}.
The mass and root-mean-square mass radius of the baryon
are examined in Sec.~\ref{MP}.
We explain the ``swelling'' mechanism of the baryon in the general
context of QCD in Sec.~\ref{SMPI}.
Through all of the sections below, by comparing the 
Skyrme model
without $\rho$ meson field and the 
brane-induced Skyrme (BIS) model
on $S^3$,
the roles of (axial) vector mesons in the baryonic matter
are discussed from a holographic point of view.
\subsection{ Energy density and field configuration profiles
\label{ECpro} }
%
\begin{figure*}[ht]
 \begin{tabular}{cc}
  \begin{minipage}{86mm}
       \resizebox{86mm}{!}{\includegraphics{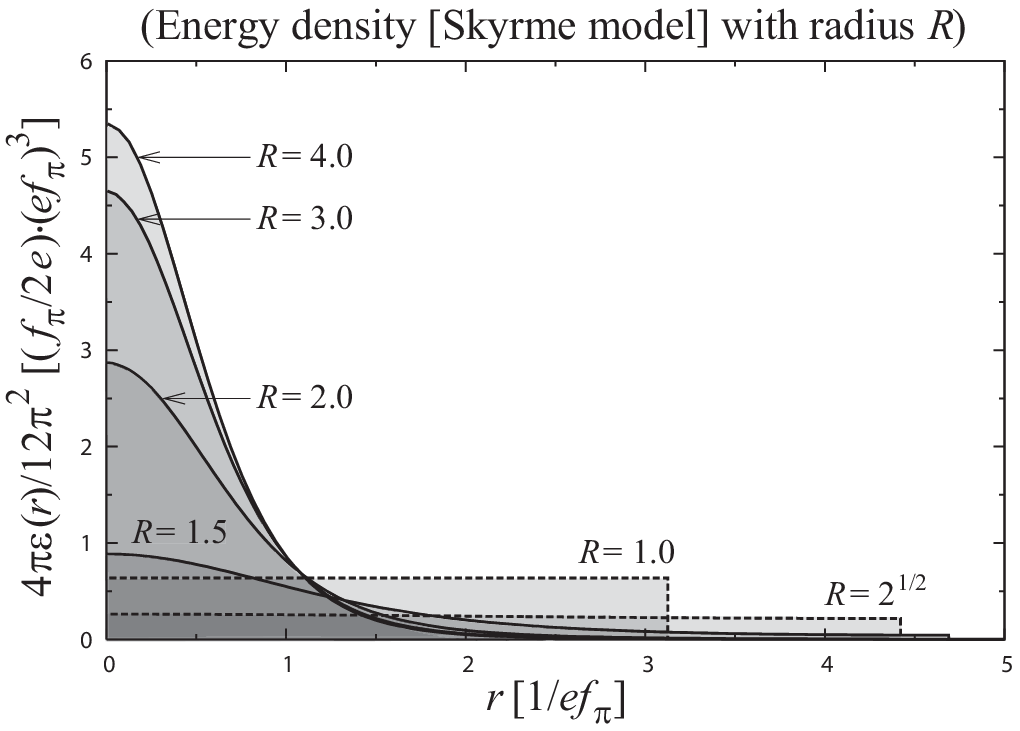}}
\caption{Energy density of single baryon for the Skyrme model
with radius $R$ of $S^3$. Below the critical radius, $R\leq R_{\rm crit}^{\rm Skyrme}=\sqrt{2}$,
energy density becomes a uniform distribution as the identity map,
shown by the dashed lines.}
  \label{fig_density1} 
\end{minipage}&
       \hspace{4mm}
  \begin{minipage}{86mm}
       \resizebox{86mm}{!}{\includegraphics{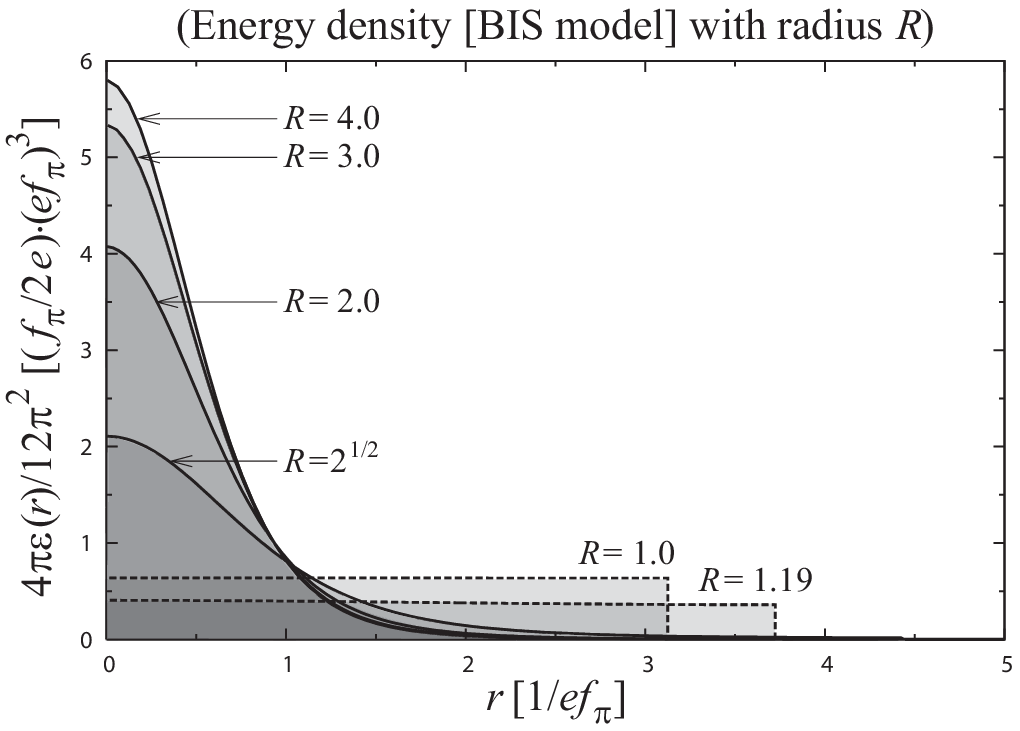}}
\caption{Energy density of single baryon for the BIS model
with radius $R$ of $S^3$. Below the critical radius, $R\leq R_{\rm crit}^{\rm BIS}=1.19$,
energy density becomes a uniform distribution as the identity map,
shown by the dashed lines.}
  \label{fig_density2}
 \end{minipage}
 \end{tabular}
\end{figure*}

In this section, we discuss the 
baryon-number density dependence of the energy density
and the field configuration profiles of 
single 
baryon for the Skyrme model and the BIS model,
by changing the size of the manifold $S^3$.
Actually,
the baryon-number density 
is represented 
as $\rho_B=1/2\pi^2R^3$
on $S^3$,
so that, as the radius $R$ of $S^3$ decreases,
the increase of total
baryon-number density
in the medium is represented in this modeling.

First, we show in Fig.\ref{fig_density1} 
the energy density of the baryon for the Skyrme model
with radius $R$ of $S^3$.
One can see that the energy density 
tends to delocalize as $R$ decreases, 
which can be regarded as some medium effects 
	in the dense baryonic matter.
Below the critical radius,
$R\leq R_{\rm crit}^{\rm Skyrme}=\sqrt{2}$,
the energy density distribution of the baryon
exactly coincides with the uniform one 
as identity map
denoted by the dashed lines in Fig.~\ref{fig_density1}. 
The transition from localized energy density into 
uniform one is 
called the ``delocalization phase transition''.

Next 
we show in Fig.\ref{fig_density2}
the same plot in the case of the BIS model.
%
The energy density of the baryon also 
tends to delocalize as $R$ decreases. 
However the delocalization 
along with the decrease of $R$
is delayed 
relative to the Skyrme model without $\rho$ meson field, 
and the energy distribution is still localized even around 
$R\sim R_{\rm crit}^{\rm Sky}=\sqrt{2}$,
which is the critical radius for the Skyrme model.
In fact, the heavy
$\rho$ meson field 
appearing
in the core region of the baryon 
is to
provide the attraction with the pion field, 
which leads
to the shrinkage of the total size of the baryon~\cite{NSK}.
Therefore,
the smaller  
radius of $S^3$, i.e., the larger baryon-number density
is needed for the BIS model to give the delocalization phase transition, 
which is discussed 
in Sec.~\ref{CDT} 
with recovering the physical units.
Then, below the critical radius, $R\leq R_{\rm crit}^{\rm BIS}=1.19$, 
the energy density distribution of the baryon in 
the BIS model becomes the uniform one
as identity map (\ref{BIS_ide}).
These
delocalization phase transitions in the Skyrmion picture
can be related with
the deconfinement of the baryon and also the chiral symmetry
restoration, 
which will be discussed 
with the order parameters in Sec.~\ref{CRDPT}.

\begin{figure}[h]
  \begin{center}
       \resizebox{89mm}{!}{\includegraphics{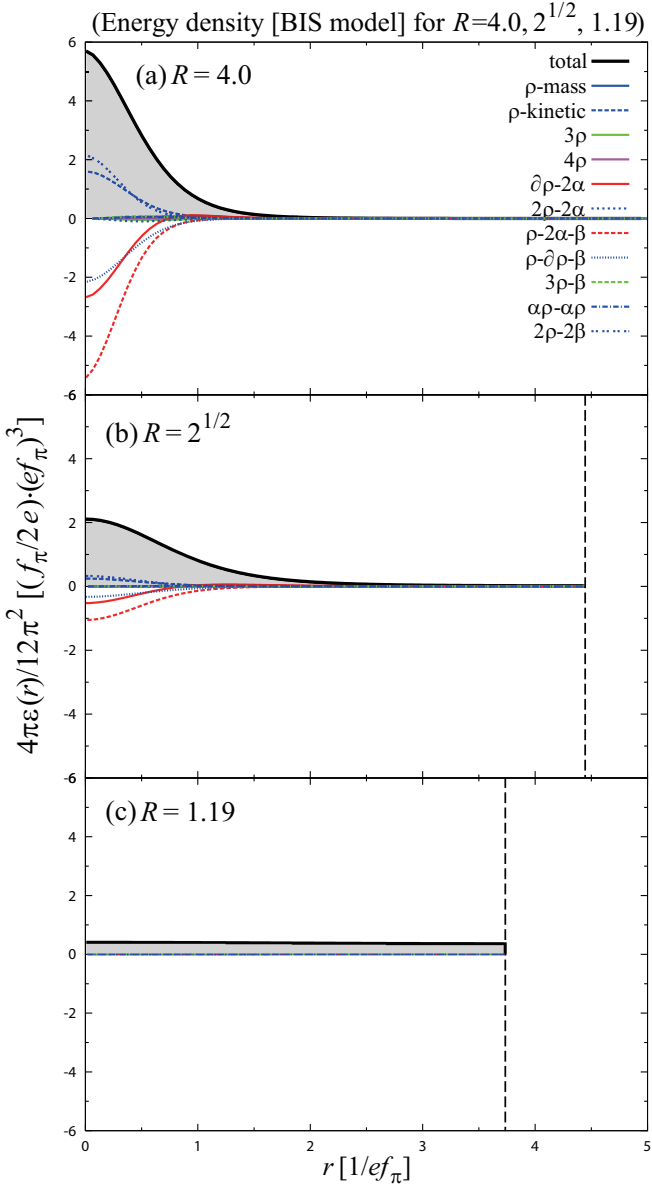}}\\
  \end{center}
\caption{Energy density of single baryon and 
$\rho$ meson contributions in the terms 
of the BIS model 
(\ref{energy_dense_Re})
for $R=4.0$, $\sqrt{2}$, and $1.19$. 
Labels for lines in (a) denote
each term in (\ref{energy_dense_Re}), e. g.
``$4\rho$'' corresponds to 
$(-1/2)g_{4\rho}\int d^4 x {\rm tr}[\rho_\mu,\rho_\nu]^2$
(See Ref.~\cite{NSK} for labels).
The vertical dashed lines in (b) and (c)
show the boundaries of $S^3$ at the south pole $r=\pi R$, respectively.
At the critical radius, $R=R_{\rm crit}^{\rm BIS}=1.19$ in (c), energy density becomes 
a uniform distribution as the identity map,
where the $\rho$ meson contributions disappear and only pion
contribution survives as the ``pion dominance''.}
  \label{fig_T_rho}
\end{figure}

We can also show in Fig.\ref{fig_T_rho}
the comparison between
the energy density of the baryon and
$\rho$ meson contributions 
in the interaction terms
of the BIS model (\ref{energy_dense_Re})
for $R=4.0$, $\sqrt{2}$, and 1.19.
For $R=4.0$ and $\sqrt{2}$,
one can see the manifest contributions from the $\rho$ meson field
in the core region of the baryon.
On the other hand, at the critical radius $R=R_{\rm crit}^{\rm BIS}=1.19$,
all the contributions from the $\rho$ meson field absolutely disappear
in the uniform phase.

\begin{figure}[ht]
  \begin{center}
\hspace*{5mm}       \resizebox{80mm}{!}{\includegraphics{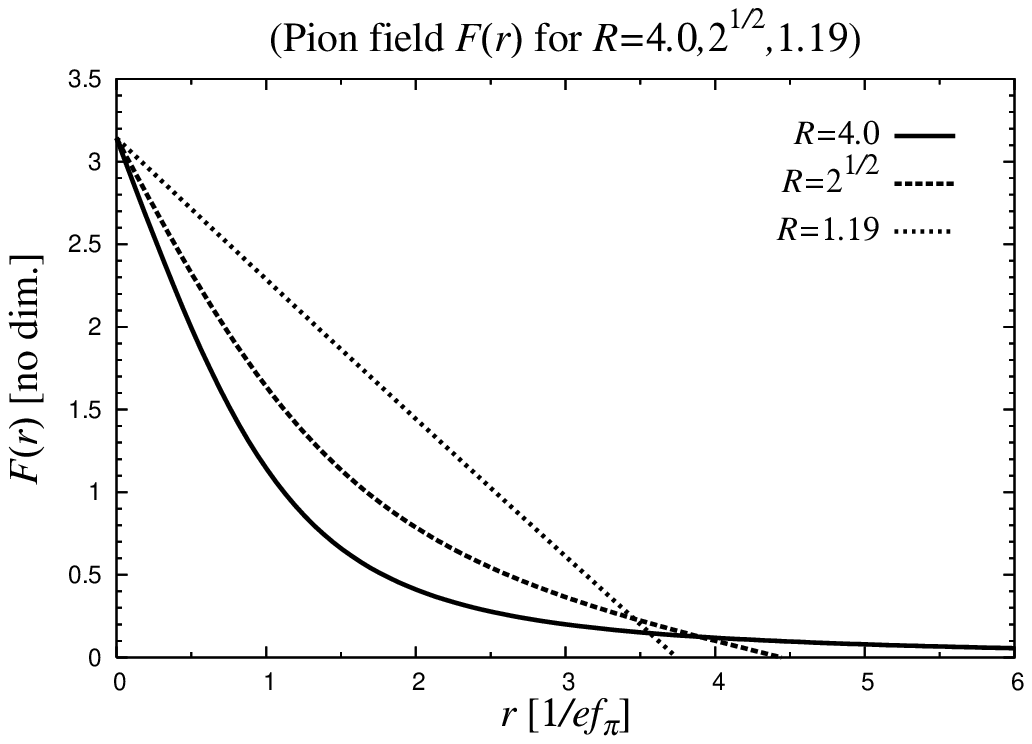}} 
  \end{center}
\caption{Pion field $F(r)$ of the BIS model for $R=4.0$, $\sqrt{2}$, and $1.19$.
At the critical radius, $R=R_{\rm crit}^{\rm BIS}=1.19$, $F(r)$ coincides with a linear
configuration as the identity map in Eq.~(\ref{BIS_ide}).}
  \label{fig_conf1}
  \begin{center}
       \vspace{5mm}
\hspace*{5mm}        \resizebox{80mm}{!}{\includegraphics{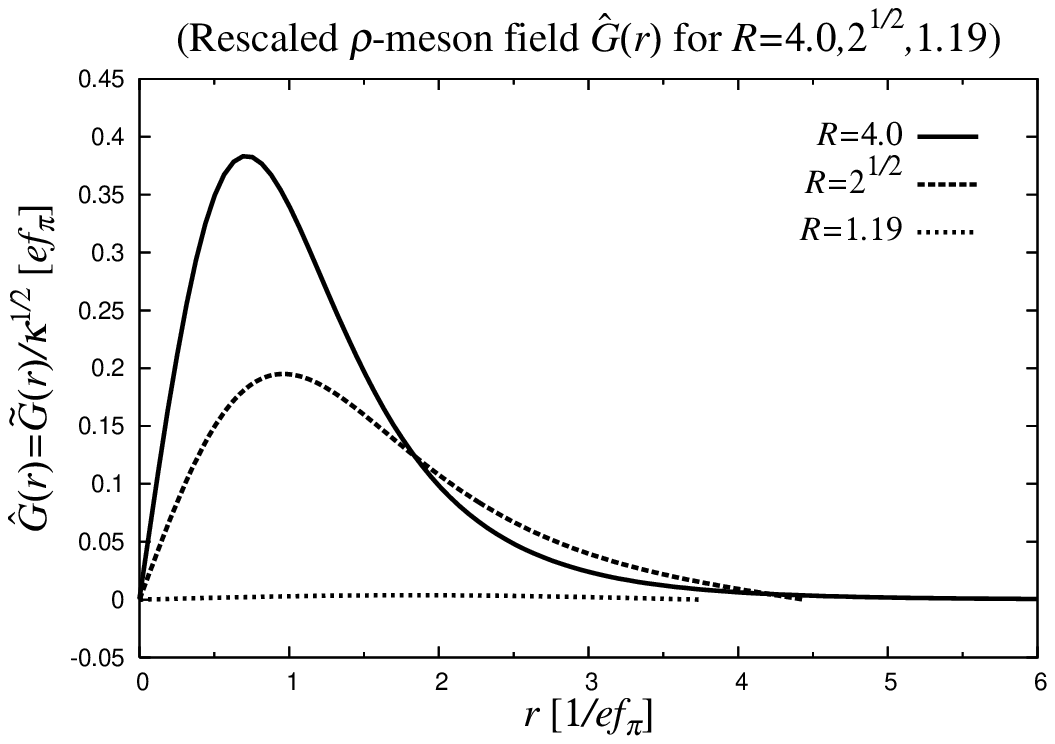}} 
  \end{center}
\caption{Rescaled $\rho$ meson field $\hat{G}(r)=\frac{\tilde{G}(r)}{\sqrt{\kappa}}$ 
of the BIS model for $R=4.0$, $\sqrt{2}$, and $1.19$.
At the critical radius, $R=R_{\rm crit}^{\rm BIS}=1.19$, $\hat{G}(r)$ becomes zero 
configuration as the identity map in Eq.~(\ref{BIS_ide}).}
  \label{fig_conf2}
\end{figure}

We also show the  
field configuration profiles of 
the pion and the $\rho$ meson in Fig.~\ref{fig_conf1}
and Fig.~\ref{fig_conf2}, respectively, in the BIS model 
for $R=4.0$, $\sqrt{2}$, and $1.19$.
As $R$ decreases,
the pion field $F(r)$ in Fig.~\ref{fig_conf1} approaches 
to the linear configuration as the identity map in 
Eq.(\ref{BIS_ide}), giving the uniform energy density distribution.
%
%
As for the $\rho$ meson field in Fig~\ref{fig_conf2},
the amplitude of $\rho$ meson field $\hat{G}(r)$
tends to decrease as $R$ decreases, 
and it absolutely disappears below the critical radius, 
$R\leq R_{\rm crit}^{\rm BIS}=1.19$ as the identity map 
in Eq.~(\ref{BIS_ide}).
These results indicate 
that the amplitude of the $\rho$ meson field decreases 
as the baryon-number density increases in the medium, 
and it absolutely disappears and only the pion field survives 
near the critical density.

\subsection{Pion dominance near critical density \label{PDNC}}
In the previous section,
we find that the $\rho$ meson field would disappear near the critical
density.
Now we propose a conjecture that 
such disappearance of the $\rho$ meson field 
near the critical density 
can be generalized to all the other (axial) vector meson fields:
$a_1, \rho', a'_{1}, \rho''\cdots$,
denoted by the field $B_\mu^{(n)}(x_\nu)$ in Eq.~(\ref{limit2_lr}) 
by the following reasons within holographic approach:
\begin{list}{}{}
\item[1)] The kinetic term of (axial) vector meson field $B_\mu^{(n)}$
on the closed manifold $S^3$ is proportional to $R^{-2}$ as
\begin{eqnarray}
{\rm tr}\ltk\partial_\mu B_\nu^{(n)}-\partial_\nu
 B_\mu^{(n)}\rtk^2\propto R^{-2},\label{kine_pro}
\end{eqnarray}
which can be derived from a simple dimensional analysis.
In general,
the kinetic energy indicates the ``kink'' energy of field
configurations.
Therefore the kinetic energy 
in Eq.(\ref{kine_pro})
suppresses the spatial dependence of the field 
configuration $B_\mu^{(n)}$ for small $R$, i.e., for high density state,
giving flat configuration in the high density phase.
\item[2)] The mass term $m_n^2 {\rm tr}\{ {B_\mu^{(n)}B_\mu^{(n)}}\}$
suppresses the absolute value of $B_\mu^{(n)}$ field
more severely in accordance with its larger mass $m_n^2$.
\item[3)] The couplings between pions and heavier (axial) vector mesons
$B_\mu^{(n)}$ with larger index $n$ 
are found to become smaller,
which is suggested by the ``bottom-up''  
dimensional deconstruction model~\cite{Son} and also
the ``top-down'' holographic approach~\cite{NSK}.
In fact, within the holographic models,
the origin of the meson mass in four dimensional space-time
can be regarded as the oscillation of meson wave function in the extra fifth dimension $z$, 
denoted by the mass relation $m_n^2=\lambda_n$ in Sec.~\ref{MEA}.
Such 
larger oscillations of heavier (axial) vector mesons wave
functions 
in the fifth dimension
have the smaller overlap with that of pions,
giving the smaller coupling constants with pions in four
dimensional space-time~\cite{NSK}.
In fact, some recent experiments with the hadron reactions
provide some interesting data, showing that the heavier (axial) 
vector mesons tend to have 
smaller width for the decay into pions 
despite of larger phase space~\cite{PDG},
which may be consistent with the prediction of holographic QCD
as their smaller coupling constants with pions as mentioned above.
Hence the effects of heavier (axial) vector mesons
should be smaller for a baryon as a large-amplitude
pion field, i.e., the chiral soliton.
\end{list}
These considerations 1), 2), 3) about the meson effective action in 
holographic QCD should support our conjecture about 
the general disappearance of
(axial) vector meson fields in the high density phase.
In other words,
{\it only the pion fields survive
near the critical density 
in the large-$N_c$ baryonic matter}.
We call this phenomena as ``pion dominance'' near the critical density.

In Sec.\ref{CDT}, we 
will 
show that,
even if the $\rho$ meson field disappears near the critical point, 
it affects the critical density of the phase transition 
through its mass and also the interactions with the pion field 
in the action $S_{\rm eff}$ in Eq.(\ref{DBI}).
In this sense, even with the pion dominance near the critical 
point proposed above,
the (axial) vector mesons 
may still affect the critical phenomena 
like the critical density
through their contributions in the effective action
if they are included.

As for the survival of the pion fields near the critical density
as the pion dominance, we give the following reason:
the unit baryon-number on the manifold $S^3$ 
comes from the boundary conditions of the 
pion hedgehog configuration $F(r)$
at the north ($r=0$) and the south ($r=\pi R$) poles on $S^3$ as 
$F(0)=\pi$ and $F(\pi R)=0$ in (\ref{bondary_S3}).
%
%
In this sense,
pions play the essential roles for the baryon-number current 
with their field boundaries.
%
Therefore the pion field cannot disappear because of the baryon-number 
constraint (\ref{bondary_S3}) for each unit cell of the manifold $S^3$.
On the other hand, there is no constraints for the other
(axial) vector meson fields, and they can
disappear in the high density phase
as some representation of ``deconfinement''.

Such dominance of the pion field near the transition point
might be somehow related with the appearance of chiral plasma modes
$\pi$ and $\sigma$ above $T_c$ of the deconfinement phase transition
observed in the lattice QCD stuty~\cite{DK}.
In fact, the screening masses were measured for a variety of
color-singlet channels with the quantum number of $\pi$, $\sigma$,
$\rho$, $b_1$, and $a_1$ mesons.
The chiral multiplet of $\pi$ and
$\sigma$ was found to appear as the bound state even above $T_c$,
whereas the other (axial) vector mesons would be in the continuum
with the threshold as the twice of the lowest Matsubara frequency, i.e.,
$2\pi T$.
Such survival of the chiral plasma modes above $T_c$
would inspire the concept of the strong-coupling quark-gluon plasma
(sQGP)
due to the long-range, non-perturbative effects of the strong
interaction
even at the highest temperature, instead of the naive free gas picture
of quarks and gluons only due to the asymptotic freedom.
Such strong correlation would also develop even at dense regime of QCD,
which may give some linking of our analysis with ``pion dominance''  
in dense QCD.

By seeing some QCD phenomenologies, one can also find that 
quark degrees of freedom are often represented by ``pions''.
For example, in the case of the chiral quark model~\cite{Nielsen,Niemi},
quark-antiquark correlations are represented
by pionic collective modes through the bosonization scheme for QCD,
giving the ``Cheshire cat picture'' where the quark dynamics are
represented by pions.
In fact, it have provided a foundation of
Skyrme soliton picture for the baryon
as a large-amplitude pion field.
These traditional phenomenologies may also support 
 the dominance of the pions near the critical density
 in the baryonic matter, 
 with the appearance of quark-gluon dynamics.

\subsection{
Mass and root-mean-square mass radius\label{MP}}
\begin{figure}[ht]
  \begin{center}
       \resizebox{80mm}{!}{\includegraphics{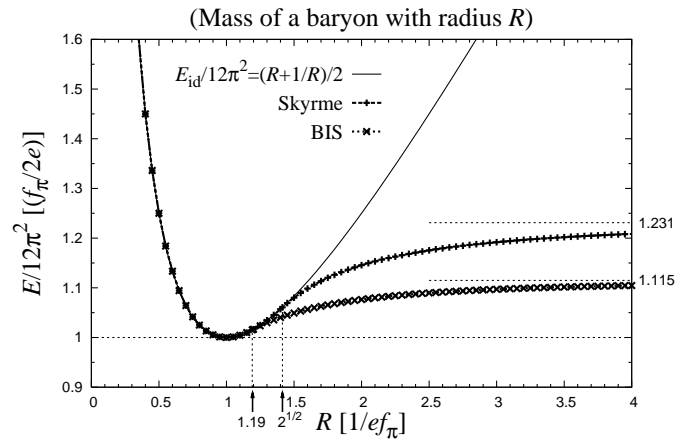}}\\
  \end{center}
\caption{Mass of single baryon for the identity map, the Skyrme model,
and the BIS model with radius $R$ of $S^3$. 
The variables ``1.231'' and ``1.115'' are the masses of the baryon on the flat
space ${\boldmath \mbox{${\rm R}$}}^3$ for the Skyrme model and the BIS model,
respectively.}
  \label{fig_masspro1}
%
\end{figure}

In this section, we discuss the baryon-number 
density dependence of the mass and 
the root-mean-square (RMS) mass radius of 
single 
baryon for the Skyrme model and the BIS model, 
by changing the size of the manifold $S^3$.

In Fig.\ref{fig_masspro1}, we show the mass of the baryon for the Skyrme
model and the BIS model with radius $R$ of $S^3$. 
%
In the case of the Skyrme model,
the mass decreases from the value on the flat space ${\boldmath \mbox{${\rm R}$}}^3$
: $1.231\times 12\pi^2[\frac{f_\pi}{2e}]$~\cite{Sky_mode} as $R$ decreases,
and it coincides with the mass of identity map below the 
critical radius, $R\leq R_{\rm crit}^{\rm Skyrme}=\sqrt{2}$~\cite{MR}.
This coincidence of numerical solution with identity map 
indicates the transition from the localized phase to the uniform phase
as the delocalization phase transition.
In the case of the BIS model,
the mass decreases from the value on ${\boldmath \mbox{${\rm R}$}}^3$
: $1.115\times 12\pi^2[\frac{f_\pi}{2e}]$~\cite{NSK} as $R$ decreases,
and it coincides with the mass of identity map below the 
critical radius, $R\leq R_{\rm crit}^{\rm BIS}=1.19$, which is smaller than 
$R_{\rm crit}^{\rm Skyrme}$
due to the shrinkage of baryon by the $\rho$ meson effects.

From a physical point of view,
the decrease of the baryon mass as $R$ decreases in Fig.~\ref{fig_masspro1}
might come from the partial 
restoration of chiral symmetry
in the dense baryonic matter, which 
will be 
discussed with the order parameter 
in Sec.~\ref{CRDPT}.
This result is related with the well-known picture of the baryon mass generation
from the quark condensate in the vacuum,
suggested in some QCD phenomenologies, e.g., 
the QCD sum rules with Ioffe formula~\cite{Ioffe}, and also other chiral
effective models~\cite{Lee, Jido}.
In fact, the decrease of the baryon mass with the chiral 
symmetry restoration
is observed in the finite temperature lattice QCD study~\cite{Rothe}.
Note also
that the decrease of the baryon mass with 
the chiral symmetry restoration in dense QCD is proposed in the framework of AdS/QCD model
as the phenomenological bottom-up constructions, called 
``hard-wall'' model~\cite{HW}.

\begin{figure}[h]
  \begin{center}
       \resizebox{80mm}{!}{\includegraphics{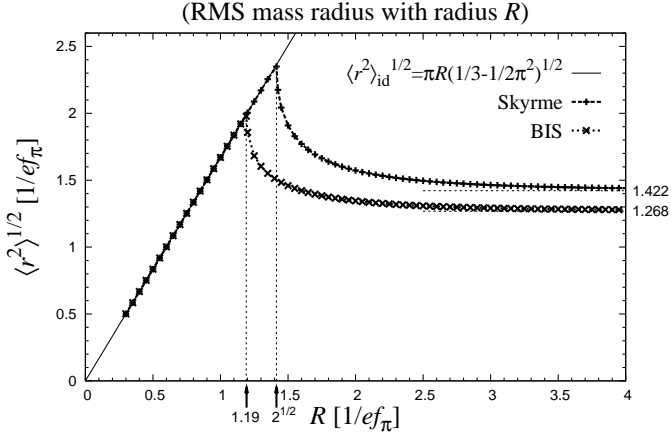}}\\
  \end{center}
\caption{RMS mass radius of single baryon for the
identity map, the Skyrme model, and the BIS
model with radius $R$ of $S^3$. 
The variables ``1.422'' and ``1.268'' are the RMS mass radius of the baryon on the flat
space ${\boldmath \mbox{${\rm R}$}}^3$ for the Skyrme model and the BIS model,
respectively.}
  \label{fig_RMS}
\end{figure}

Next we analyze of the RMS mass radius of the baryon
for the Skyrme model and the BIS model.
By using the normalized energy density $\bar{\varepsilon}(r)\equiv
\varepsilon (r)/E$ ($\varepsilon (r)$ is total energy density and $E$ is
the mass of single baryon),
the RMS mass radius can be naturally introduced 
on $S^3$ with radius $R$ as
\begin{eqnarray}
\sqrt{\langle r^2 \rangle}
&=&\ldk\int_{S^3}d^3x\cdot \bar{\varepsilon}(r) r^2 \rdk^{1/2}\nonumber\\
&\equiv& \ldk \int_0^{\pi R} 
4\pi dr R^2\sin^2\frac{r}{R}\cdot\bar{\varepsilon}(r) r^2
\rdk^{1/2},\label{RMS_1}
\end{eqnarray}
where $R^2\sin^2\frac{r}{R}$ denotes the measure of the curved manifold $S^3$.
In the case of the identity map,
the RMS mass radius can be analytically calculated
by using the normalized uniform energy density for the identity map : 
$\bar{\varepsilon}_{\rm id}\equiv 1/2\pi^2 R^3$ as 
\begin{eqnarray}
\sqrt{\langle r^2 \rangle_{\rm id}} &\equiv& 
\ldk \int_0^{\pi R} 
4\pi dr R^2\sin^2\frac{r}{R}\cdot\bar{\varepsilon}_{\rm id} r^2
\rdk^{1/2} \nonumber\\
&=& \pi R\sqrt{\frac{1}{3}-\frac{1}{2\pi^2}}. \label{RMS_ide}
\end{eqnarray}

In Fig.\ref{fig_RMS},
we show the RMS mass radius for the Skyrme model and the BIS model
with radius $R$ of $S^3$.
One can find that the
RMS mass radius in each model increases non-linearly around its critical radius
as $R$ decreases, indicating the ``swelling'' phenomenon of the baryon in 
high density phase.
Such swelling of the baryon
can also be seen 
in the analysis
of the Skyrme crystal on the three-dimensional cubic lattice~\cite{Kle}
and also 
the finite density bag model~\cite{Ichie}.
After the delocalization phase transition,
the energy density of single baryon
is uniformly saturated over the surface of $S^3$.
Then the RMS mass radius decreases linearly with 
the decrease of $R$ as 
seen 
in Eq.(\ref{RMS_ide}).
(In the uniform phase, baryons are no longer localized, 
and the RMS mass radius may have less physical meaning.)

\subsection{Swelling mechanism and its phenomenological implications \label{SMPI}}
\begin{figure*}[t]
  \begin{center}
   \hspace*{0mm}
       \resizebox{160mm}{!}{\includegraphics{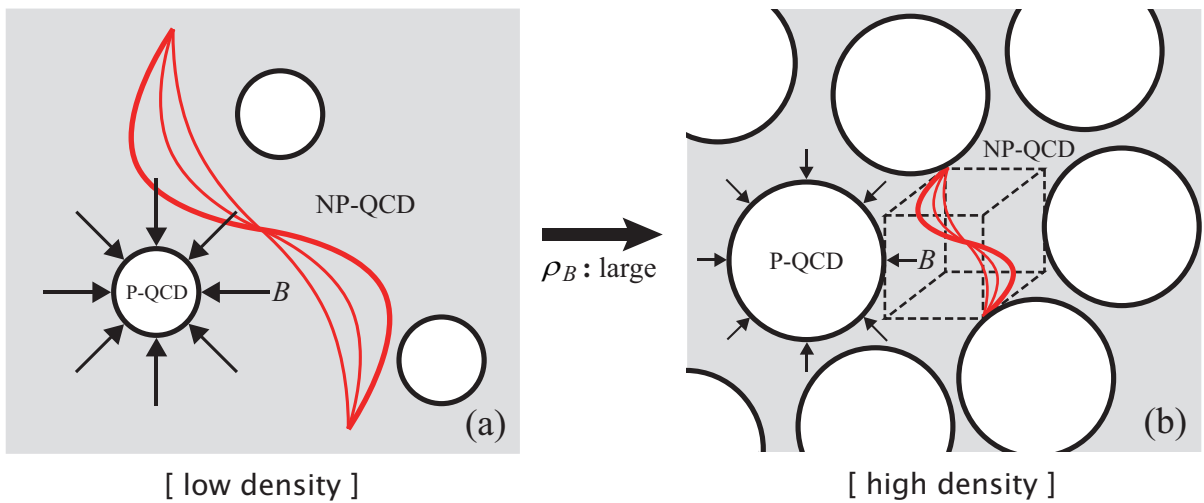}}\\
  \end{center}
\caption{Schematic figure of the finite density bag model
for low density case (a) and high density case (b):
the baryon is represented by a ``bag'' with 
perturbative QCD vacuum (P-QCD) surrounded by
Dirichlet boundaries, supported by 
the non-perturbative QCD vacuum (NP-QCD) with
the bag pressure $B$.
Several waves in (a) and (b) represent the non-perturbative effects of QCD
with certain wave length.
A cube with dashed lines in (b) imitates a definite region of
non-perturbative QCD vacuum
surrounded by the surfaces of the bags with Dirichlet boundaries,
to discuss the mechanism of ``swelling'' in the high density baryonic matter.}
  \label{fig_fbag}
\end{figure*}
In the previous section, we numerically
find the swelling of the baryon by taking the
Skyrmion picture 
for the baryon analysis.
%
%
Actually, as discussed in Sec.~\ref{BIS_R3},
the Skyrmion picture is naturally derived from QCD 
through the new concept of holography.
%
Therefore
one can expect the 
swelling phenomenon in QCD itself.
In this section, we try to 
give a qualitative explanation about the
swelling mechanism of a baryon
in terms of QCD, by especially 
referring to 
contexts of
the lattice QCD~\cite{Rothe} and 
also the finite density bag picture~\cite{Ichie}.

First, in lattice QCD study,
the non-perturbative QCD vacuum is analyzed within a finite box of 
periodic boundaries (and also Dirichlet boundaries)
in the Euclidean space.
The 
non-perturbative effects 
with long wave length tends to be ``blocked out''
from such finite box of definite boundaries,
so that the QCD vacuum in
the finite box
is found to approach
to the perturbative one.
Also, the finite-temperature phase transition 
in lattice QCD
occurs because of the definite boundaries along the
imaginary time axis in Euclidean space;
as the temperature increases, the QCD vacuum in the Euclidean space
is more closely packed
along the imaginary time axis to block out the non-perturbative effects
with long wave length, which eventually gives the transition
into the perturbative QCD vacuum, like 
the chiral symmetry restoration
and also the deconfinement.

Next we 
combine the above picture into
the finite density bag model,
to explain the mechanism of the swelling in QCD.
In the  
bag model, the baryon is represented 
by a ``bag'' with the perturbative QCD vacuum surrounded by
the Dirichlet boundaries, 
which is 
supported by the non-perturbative QCD vacuum 
with the bag pressure as 
shown in Fig.~\ref{fig_fbag}(a).
As 
a number of bags 
increases, 
representing the high density baryonic matter,
the non-perturbative QCD vacuum around the bags
is more closely packed 
within a definite region,
imitated by a cube with dashed lines in Fig.~\ref{fig_fbag}(b).
Then,
in analogy with the case of lattice QCD study discussed above,
the QCD vacuum around the bags 
should approach
to the perturbative one by blocking out the
non-perturbative effects with long wave length as
shown by the several waves in Fig.~\ref{fig_fbag}.
Therefore the bag pressure as a non-perturbative effect
decreases
as the number of bags increases,
so that it eventually gives the swelling of bags.
Such swelling in turn decreases the
region of
non-perturbative QCD vacuum around the bags
to give the decrease of the bag pressure again.
In this sense, such swelling could occur {\it non-linearly}
as shown 
in Fig.~\ref{fig_RMS}, giving some drastic change of features like 
``deconfinement'' in the baryonic matter.
One can expect some resemblance between the finite density bag model
and the Skyrme matter in our study 
with respect to baryonic matter, so that
the swelling phenomena 
as shown in Fig.~\ref{fig_RMS}
could be understood on the same footing
provided above.

Now
we also suggest the physical effects of the swelling in the high density
baryonic matter.
Here we take into account possible baryon excitation 
with the chiral soliton picture.
Since
a moment of inertia of the baryon is
related with the size of the baryon,
some non-linear increase in the moment 
of inertia around critical density is
expected along with the swelling of a baryon.
%
%
In the semiclassical quantization procedure of a Skyrme soliton,
the baryon mass spectra is given~\cite{KFLiu} as
\begin{eqnarray}
M_J=M_{\rm HH}+\frac{J(J+1)}{2{\cal I}},\label{SQP}
\end{eqnarray}
where $M_{\rm HH}$ is static hedgehog mass and ${\cal I}$ is the moment
of inertia of a baryon.
The quantum number $J$ in Eq.~(\ref{SQP}) denotes the spin $S$
and the isospin $I$ of the baryon as 
$J(=S=I)=\frac{1}{2}$ for $N$ and 
$J(=S=I)=\frac{3}{2}$ for $\Delta$.
The baryon mass splitting in the second term in Eq.(\ref{SQP})
is proportional to the inverse of the moment of inertia,
indicating that a larger object cannot be easily rotated.
Therefore we can now propose that
the baryon mass splitting decreases
as the density increases in the medium, because 
the moment of inertia ${\cal I}$ increases due to the swelling.
In other words,
the {\it individuality} of each baryon like
$N$, $\Delta$, etc., would be
{\it lost} at least with respect to their mass spectra
as the density increases.
Such lost of individuality
may be regarded as some precursory phenomena of
deconfinement,
where the dominant degrees of freedom in 
the hadronic matter
shifts from the confined composites as baryons (and mesons)
into quarks and gluons.

Such swelling of the baryon leads to some interesting
phenomenological realizations.
As one possibility, here
we propose the stable 
$N$-$\Delta$ mixed matter in dense QCD.
For simplicity, we consider 
the ``symmetric nuclear matter'', where only
the strong interaction 
is taken into account without
the electro-magnetic interaction.
In the low density case, there exists
large $N$-$\Delta$ splitting as $(M_\Delta-M_N)\sim 290{\rm MeV}$,
and the Fermi surface of the nucleons $\mu_N$ may locate below 
the threshold of $\Delta$ isobars, i.e., its mass $M_\Delta$.
This situation
gives only the nucleon degrees of freedom as nuclear matter.
Now, if the swelling occurs as the density increases, 
the $N$-$\Delta$ splitting 
should decrease as discussed in Eq.~(\ref{SQP}).
Then, the Fermi surface of the nucleons $\mu_N$
would excess $M_\Delta$
to give the $\Delta$ isobar degrees of freedom in the medium
through the equilibrium process $N+\pi\leftrightarrow\Delta$.
In this sense,
$N$-$\Delta$ mixed matter would be realized
in dense QCD 
because of the swelling of a baryon.
Such $N$-$\Delta$ mixed matter may appear
in the deep
interior of neutron stars between the nuclear 
crust and the core of quark matter as the precursor of deconfinement, giving 
some softening of the EOS of neutron stars
relative to the analysis without the mixed matter.
While 
these are qualitative discussions,
we feel that these explanations catch some essential aspects
of baryonic matter with baryon excitation in view of its ``swelling'' nature.
 
\section{Phase transitions with order parameters
\label{CRDPT}}
%
In this section we discuss 
the delocalization phase transition
in view of the deconfinement,
and also the chiral symmetry restoration 
by introducing the proper order parameters in the Skyrme model and
the BIS model.
The relations between these phase transitions are also
considered by referring to other QCD phenomenologies in the end of Sec.~\ref{CSR}.

\subsection{Delocalization phase transition}

First we consider 
the delocalization phase transition
with the order parameter.
By using the normalized energy density $\bar{\varepsilon}({\bf x})\equiv
\varepsilon ({\bf x})/E$ ($\varepsilon ({\bf x})$ is
 total energy density and $E$ is
the mass of single Skyrmion),
one can introduce the spatial fluctuation $\Phi(R)$
of the energy density around the uniform energy density 
distribution~\cite{Jackson} as
\begin{eqnarray}
\Phi(R)&\equiv\frac{1}{2}\int_{S^3}d^3 x |\bar{\varepsilon}({\bf x})
                                     -\bar{\varepsilon}_{\rm id}|,\label{LOP1}
\end{eqnarray}
where $\bar{\varepsilon}_{\rm id}\equiv\frac{1}{2\pi^2 R^3}$
is the normalized uniform energy density for the identity map.
If the Skyrmion is well localized like the delta function,
the energy density $\bar{\varepsilon}({\bf x})$
is almost 
decoupled with uniform energy density distribution $\bar{\varepsilon}_{\rm id}$
through the spatial integral over $S^3$.
Therefore its spatial fluctuation $\Phi(R)$ becomes,
\begin{eqnarray}
\Phi(R)&\sim\frac{1}{2}\int_{S^3}d^3 x \ltk|\bar{\varepsilon}({\bf x})|+
                                     |\bar{\varepsilon}_{\rm id}|\rtk=1.\label{LOP2}
\end{eqnarray}
On the other hand, if the Skyrmion gives uniform energy density
distribution
as the identity map,
the energy density $\bar{\varepsilon}({\bf x})$
coincides with $\bar{\varepsilon}_{\rm id}$,
giving
\begin{eqnarray}
\Phi(R)=0.
\end{eqnarray}
In this sense, $\Phi(R)$ in Eq.(\ref{LOP1}) 
measures the amount of localization in the energy density distribution
of the Skyrmion, regarded as the order parameter of
delocalization phase transition in the baryonic matter.

\begin{figure}[t]
  \begin{center}
       \resizebox{86mm}{!}{\includegraphics{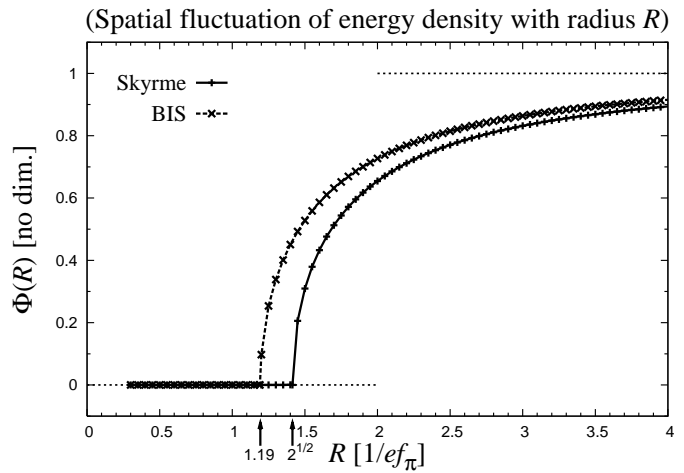}}\\
  \end{center}
\caption{Spatial fluctuation $\Phi(R)$ of the energy density 
of single baryon for the Skyrme model and the
 BIS model with radius $R$ of $S^3$.}
  \label{fig_DF}
\end{figure}

In Fig.~\ref{fig_DF}, 
we show the value of $\Phi(R)$
for the Skyrme model and the BIS model 
with radius $R$ of $S^3$.
For sufficiently large $R$
as the low density state, the Skyrme soliton is 
well
localized for both the Skyrme model and the BIS model,
giving $\Phi(R)\sim 1$.
With the decrease of $R$, 
the baryon tends to delocalize 
with the decrease of $\Phi(R)$,
regarded 
as some medium effect in the 
baryonic matter.
The delocalization phase transition 
into the uniform phase with $\Phi(R)=0$ occurs
at the critical radius $R=R_{\rm crit}^{\rm Skyrme}=\sqrt{2}$ 
for the Skyrme model
and at $R=R_{\rm crit}^{\rm BIS}=1.19$ for the BIS model.
Actually, the delocalization phase transition in the BIS model is delayed
along with the decrease of $R$
because of the heavy $\rho$ mesons in the core region of the baryon,
discussed in Sec.~\ref{ECpro}.

Here one may have some temptation to relate
the delocalization phase transition
with the ``deconfinement'' of the baryon in QCD,
which should be carefully discussed.
Actually, there 
have been
several conflicts 
about the meaning of the delocalization phase transition
in the standard Skyrme model 
with finite density~\cite{Kle,MR,Rein,Jackson,Forkel}.
Within the traditional works in 1970',
the direct linking of the Skyrme model with
QCD was still uncertain.
Furthermore,
the standard
Skyrme model does not manifestly
include
quarks and gluons,
so that large care should be taken to relate the delocalization of
the Skyrmion with the deconfinement of the baryon 
as the appearance of 
quark-gluon dynamics.
In our case, however, the new concept of holography
 allows us to derive
the Skyrmion picture
from QCD itself, 
which is discussed in Sec.~\ref{BIS_R3}.
Moreover, 
one should also note that 
quark degrees of freedom can
be represented by ``pions'', 
suggested in some QCD phenomenologies
with ``Cheshire cat picture'' as in the chiral quark model~\cite{Nielsen,Niemi}
derived from the bosonization scheme for QCD.
With these backgrounds, here we
propose that the delocalization phase transition
in the BIS model can be 
more
admissibly related 
with
the deconfinement of the baryons in QCD,
relative to the standard Skyrme models in 1970'.

\subsection{Chiral symmetry restoration in non-linear realization\label{CSR}}
Next we consider the chiral symmetry restoration
with the order parameter.
Normally, in case of the linear sigma model,
the chiral symmetry restoration 
can be signaled by the vanishing of the expectation value
of the meson fields
at the matter ground state as
\begin{eqnarray}
\langle
\sigma(x_\mu)^2+{\bf \Pi}(x_\mu)^2
\rangle^*
=0,\label{NCR1}
\end{eqnarray}
where the sigma meson field $\sigma(x_\mu)$ and 
three pseudoscalar pion fields ${\bf \Pi}(x_\mu)$
are called
as the ``linear'' realization of chiral symmetry.
We relate this terminology to 
the non-linear sigma model,
since the meson effective action and also the Skyrmion picture
are based on the
``non-linear'' realization of the chiral symmetry.
In the non-linear sigma model, the meson fields 
$\sigma(x_\mu)$ and ${\bf \Pi}(x_\mu)$
are introduced from the chiral
field $U(x_\mu)\in {\rm SU}(N_f)_{\rm A}$ as 
\begin{eqnarray}
U(x_\mu)&=&e^{i\tau_a\pi_a(x_\mu)/f_\pi}\nonumber\\
        &=&\cos\{|\pi|/f_\pi\}+i\tau_a\hat{\pi}_a\sin\{|\pi|/f_\pi\}\nonumber\\
        &\equiv& 
	\{ \sigma(x_\mu)+i\mbox{\boldmath
	 $\tau$}\cdot{\bf\Pi}(x_\mu)\}/f_\pi.\nonumber\\
          &&    (|\pi|\equiv\sqrt{\pi_a\pi_a},
	 \hat{\pi}_a\equiv\pi_a/|\pi|) \label{SP1}
\end{eqnarray}
Since $U(x_\mu)$ is unitary 
matrix with the condition
$U^{\dagger}(x_\mu)U(x_\mu)=1$, 
the squared sum of the meson fields 
$\sigma(x_\mu)$
and ${\bf \Pi}(x_\mu)$ in Eq.~(\ref{SP1})
should be constant everywhere as 
\begin{eqnarray}
\sigma(x_\mu)^2
+{\bf \Pi}(x_\mu)^2=f_\pi^2.\label{NLS1}
\end{eqnarray}
Therefore, as far as the action is written in the chiral field
$U(x_\mu)$ with fixed $f_{\pi}$,
the meson fields
$\sigma(x_\mu)$
and ${\bf \Pi}(x_\mu)$ are forced on a surface of a
three-dimensional closed manifold $S_{\rm int}^3$ with finite radius
$f_\pi$ in the internal space.
In fact, the existence of such closed manifold $S_{\rm int}^3$
is essential 
for the concept of
a Skyrmion,
which is a non-trivial winding of the compactified physical manifold 
$S^3$ around the other closed manifold $S_{\rm int}^3$ 
with conserved topological charge, belonging to the homotopical classification
$\pi_3(S^3)={\bf Z}$~\cite{Steenrod}.
However, by comparing Eq.~(\ref{NLS1}) with Eq.~(\ref{NCR1}),
one would always encounter a problem of how to describe
the chiral symmetry restoration with its non-linear realization.

Now in this paper,
we take a spatially-averaged condensate of
the meson fields $\sigma(x_\mu)$ and ${\bf \Pi}(x_\mu)$ 
as the order parameter of the chiral symmetry restoration
with the non-linear realization~\cite{Forkel} as
\begin{eqnarray}
\{\langle \sigma(x_\mu) \rangle^2+
                    \langle{\bf \Pi}(x_\mu)\rangle^2\}/f_\pi^2.\label{EXP1}
\end{eqnarray}
Here,  
the bracket in Eq.(\ref{EXP1}) denotes
the three-dimensional spatial average in the medium with volume $V$ as 
\begin{eqnarray}
\langle \Theta(x_\mu)\rangle \equiv \frac{\int_V d^3 x \Theta(x_\mu)}
                                         {\int_V d^3 x }.\label{Ave1}
\end{eqnarray}
Such spatially-averaged condensate of the meson fields is somehow 
similar to the spatially-averaged magnetization 
$\langle {\bf M}({\bf x}) \rangle$
as the global order parameter of the ferromagnetic material like bulk
iron.
Within the bulk iron at 
sufficiently high temperature,
the spins of the ions orient randomly to get the entropy gain
in the free energy.
As the temperature decreases below the critical temperature
without the external magnetic field,
there appear magnetic domains with non-zero local magnetization 
${\bf M}({\bf x})\neq 0$.
However, because of the total angular momentum conservation,
the spatially-averaged magnetization $\langle {\bf M}({\bf x}) \rangle$
should vanish with the appearance of a complex structure
of magnetic domain walls within the bulk iron,
which gives {\it no} breaking of the global spatial
symmetry macroscopically in the bulk
material as $\langle {\bf M}({\bf x}) \rangle=0$.
In this sense, the spatially-averaged condensate of the meson fields
in Eq.(\ref{EXP1})
can be regarded as the ``global''
order parameter of the chiral symmetry in
the bulk hadronic matter.

By taking the hedgehog configuration Ansatz 
$\pi_a(x_\mu)/f_\pi=\hat{x}_a F(r)$ as in Eq.(\ref{HH}),
the meson fields $\sigma(x_\mu)$ and ${\bf \Pi}(x_\mu)$ can be written as
\begin{eqnarray}
\sigma(x_\mu)/f_\pi
&=&\cos\{\pi(x_\mu)/f_\pi\}=\cos F(r),\label{HH_sigma}\\
{\bf \Pi}(x_\mu)/f_\pi&=&\hat{\mbox{{\boldmath \mbox{${\rm \pi}$}}}}\sin\{\pi(x_\mu)/f_\pi\}
                       =\hat{\mbox{{\boldmath \mbox{${\rm x}$}}}}\sin F(r).\label{HH_pi}
\end{eqnarray}
Note here that the pion field
with the hedgehog ansatz in Eq.(\ref{HH_pi})
is proportional to the unit directional vector 
$\hat{\mbox{{\boldmath \mbox{${\rm x}$}}}}$,
so that its spatial average 
becomes trivially zero.
Therefore
only the sigma meson field 
should be considered as the 
global
order parameter of the chiral
symmetry.
If a Skyrmion is well 
localized around the north pole at $r=0$ on
the manifold $S^3$,
the classical meson filed configuration $F(r)$ becomes zero
almost everywhere except for the localized point $r\sim 0$. 
Therefore 
the sigma meson configuration in Eq.(\ref{HH_sigma}) becomes
$\sigma(x_\mu)/f_\pi\sim 1$ 
almost everywhere.
In other words, the vacuum is so much oriented 
even globally in the medium,
thus its spatial average becomes non-zero as
\begin{eqnarray}
\langle\sigma(x_\mu)\rangle/f_\pi\sim 1,\label{OP1}
\end{eqnarray}
indicating the appearance of  
the chiral symmetry broken phase.
On the other hand, if the Skyrmion gives uniform energy density distribution
as the identity map $F(r)=\pi-r/R$ 
in Eq.(\ref{BIS_ide}),
the sigma meson configuration becomes
$\sigma(x_\mu)/f_\pi=\cos\frac{r}{R}$,
changing monotonously
from $-1$
to $1$ over the coordinate space $S^3$ through the arc distance 
$r\in [0,\pi R]$.
Therefore its spatial average vanishes as
\begin{eqnarray} 
\langle\sigma(x_\mu)\rangle/f_\pi=0,\label{DOP1}
\end{eqnarray}
indicating the appearance of the chiral symmetry restored phase.
These considerations about
Eqs.~(\ref{OP1}) and (\ref{DOP1}) also indicate that the 
chiral symmetry restoration is indirectly related with
the energy density distribution 
in the hadronic matter
through the classical meson field configurations. 

\begin{figure}[h]
  \begin{center}
       \resizebox{86mm}{!}{\includegraphics{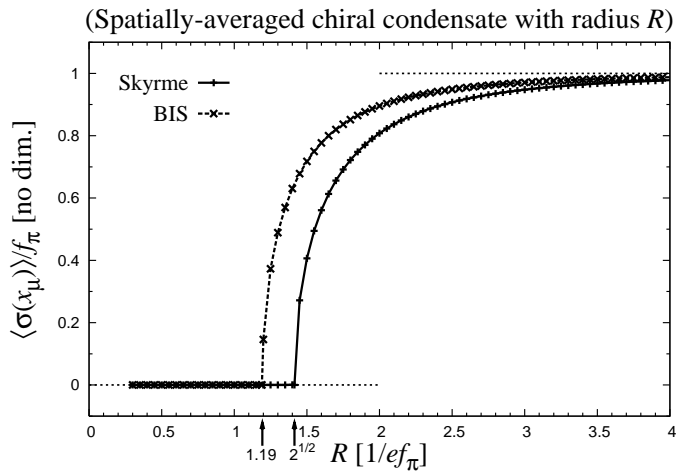}}\\
  \end{center}
\caption{Spatially-averaged condensate 
of the sigma meson field $\langle\sigma(x_\mu)\rangle/f_\pi$
of single baryon for the Skyrme model and the BIS model
with radius $R$ of $S^3$.}
  \label{fig_CC}
\end{figure}

In Fig.~\ref{fig_CC}, we show 
the spatially-averaged condensate  
$\langle\sigma(x_\mu)\rangle/f_\pi$
for the Skyrme model and the BIS model
with radius $R$ of $S^3$.
The spatial average with
the hedgehog configuration (\ref{HH_sigma})
is explicitly taken for each closed manifold $S^3$ as
\begin{eqnarray}
\langle\sigma(x_\mu)\rangle/f_\pi&=&\langle \cos F(r)\rangle\nonumber\\
                           &=&\frac{\int_{S^3}d^3x 
                              \cos F(r)}{\int_{S^3}d^3x}\nonumber\\
                           &=&\frac{\int_{0}^{\pi
			    R}4\pi dr R^2\sin^2\frac{r}{R}\cdot \cos F(r)}  
                             {2\pi^2 R^3}. \label{CCS3}
\end{eqnarray}
For sufficiently large $R$ 
as the low density state, 
the Skyrme soliton is well localized for both 
the Skyrme model and the BIS model,
as previously shown in Fig~\ref{fig_DF}.
Therefore, as discussed in Eq.~(\ref{OP1}),
the spatially-averaged 
condensate of the meson field becomes 
$\langle\sigma(x_\mu)\rangle/f_\pi\sim 1$
as the chiral
symmetry breaking in both models as shown in Fig.~\ref{fig_CC}.
In this sense, $\rho$ meson field (and also the other vector 
and axial vector meson fields, if they are included) would have 
less importance for the chiral symmetry breaking in the bulk hadronic matter
as far as sufficiently localized baryon appears in the medium.
With the decrease of $R$,
$\langle\sigma(x_\mu)\rangle/f_\pi$
monotonously decreases,
which corresponds to the partial chiral symmetry restoration
as the medium effect in the baryonic matter.
The chiral symmetry is fully restored
with $\langle\sigma(x_\mu)\rangle/f_\pi=0$
at the critical radius $R=R_{\rm crit}^{\rm Skyrme}=\sqrt{2}$ 
for the Skyrme model
and at $R=R_{\rm crit}^{\rm BIS}=1.19$ for the BIS model.
Actually the small radius $R$, i.e.,
the larger baryon-number density
is needed 
to give the chiral symmetry restoration due to 
the $\rho$ meson fields,
in similar to the case of the delocalization phase transition
as previously discussed.

Finally, we comment about the relation
between
the deconfinement and 
the chiral symmetry restoration in QCD,
which
has been discussed 
so long in these decades of the 
hadron physics.
The relation between two phase transitions 
is nontrivial since they are characterized by 
the different symmetries:
the global chiral symmetry 
${\rm SU}(N_f)_{\rm L}\times{\rm SU}(N_f)_{\rm R}$,
and Z(3) symmetry as the center of ${\rm SU}(3)_c$
gauge group.
Note that the former can be defined in the massless quark limit,
while the latter in the heavy quark limit,
leading to the difficulty to discuss their relation.

Despite of the difficulty, several implications have been obtained:
In finite temperature, the lattice QCD studies
suggest that
the two phase transitions  
occur at
the same critical temperature $T_c\sim 170{\rm MeV}$~\cite{Rothe}.
Such simultaneous occurrence is 
supported by the analyses
of the Nambu-Jona-Lasinio model with a Polyakov loop (PNJL model)
as a low-energy effective theory of QCD~\cite{Weise}.
%
%
Actually,
these two phase transitions are separated with respect to
their dominant symmetries as mentioned above, so that 
such mysterious coherence between two phase transitions
is often compared to some kinds of ``entanglement''~\cite{Weise}.
Furthermore,
the simultaneous occurrence of the two phase transitions
is also suggested by
the recent analysis of the holographic QCD
with D4/D8/$\overline{\rm D8}$ 
multi-D brane configurations:
two independent chirality spaces on the D8 and $\overline{\rm D8}$ 
branes respectively are connected with each other 
(chiral symmetry breaking)
by the ``worm hole'' of the D4 supergravity background
into which the colored objects like quarks and gluons are absorbed 
(color confinement)~\cite{NSK}.
In this sense,
these two phase transitions are more
directly related with each other
as just single event on the ``worm hole'' in the extra-dimensions,
the effects of which might appear as some kinds of mysterious
entanglement in view of the four-dimensional space-time.
%

Now, the situation has not been clear in finite density.
In this work, we indirectly relate the two phase transitions 
through the meson field configurations.
Admittedly our approach includes only meson fields 
and baryons appear as mesonic solitons. However,
if the Cheshire cat picture holds
and quark-gluon dynamics can be indirectly expressed 
in the meson dynamics 
as an effective model,
our results could be interpreted as the simultaneous occurrence of 
two phase transition in the finite density QCD.
Investigations in this direction are interesting and
should be developed further.

\section{Critical density with physical units \label{CDT}}

In this section, we 
show
the critical densities of the phase transitions
for the Skyrme model and the BIS model with recovering the physical units.

In this study, single baryon is placed on a closed manifold $S^3$ with
the surface volume
$2\pi^2 R^3$,
so that the total baryon-number density $\rho_B$ can be given as
\begin{eqnarray}
\rho_B=\lk 2\pi^2 R^3 \rk^{-1} \lk ef_\pi \rk^3
 \ldk{\rm fm}^{-3}\rdk. \label{rho_cri1}
\end{eqnarray}

\begin{figure}[t]
  \begin{center}
       \resizebox{86mm}{!}{\includegraphics{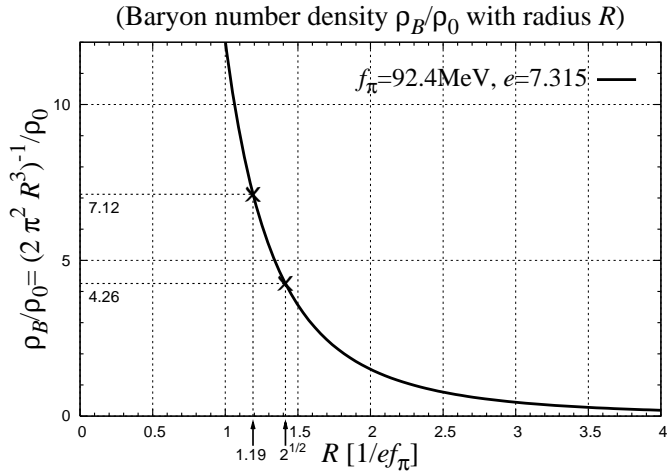}}\\
  \end{center}
\caption{Total baryon-number density $\rho_B$
in Eq.~(\ref{rho_cri1})
with radius $R$ of $S^3$,
where the experimental inputs are taken as $f_\pi=92.4{\rm MeV}$ 
and $e=7.315$ in 
Eqs.~(\ref{exp_input1}) and (\ref{e_sky_exp_input}).
The vertical axis is given as $\rho_B$ 
divided by 
the normal nuclear density $\rho_0\simeq 0.17 {\rm fm}^{-3}$.
Critical densities of the phase transitions
for the Skyrme model and the BIS model are also shown 
for each critical radius as
$R_{\rm crit}^{\rm Skyrme}=\sqrt{2}$ and $R_{\rm crit}^{\rm BIS}=1.19$.}
  \label{fig_rho_c}
\end{figure}

Now, in the holographic model,
by fixing two parameters, e.g.,
experimental inputs for $f_\pi$ and $m_\rho$,
all the physical quantities like
masses and the coupling constants are uniquely determined, which
is a remarkable consequence of the holographic approach
discussed in Sec.~\ref{MEA}.
First, we take $f_\pi$ and $m_\rho$ as experimental values,
\begin{eqnarray}
f_\pi=92.4 {\rm MeV},\hspace{4mm} m_\rho=776.0 {\rm MeV}. \label{exp_input1}
\end{eqnarray}
Then the Skyrme parameter $e$ in Eq.(\ref{e_kappa})
can be uniquely determined as 
\begin{eqnarray}
e\simeq 7.315.  \label{e_sky_exp_input}
\end{eqnarray}
With these experimental inputs,
we show in Fig.~\ref{fig_rho_c}
the total baryon-number density $\rho_B$
in Eq.~(\ref{rho_cri1})
as a function of radius $R$ of $S^3$,
divided by the normal nuclear density $\rho_0\simeq 0.17 {\rm fm}^{-3}$. 
From Fig.\ref{fig_rho_c} and also Eq.(\ref{rho_cri1}),
the critical densities for the Skyrme model ($R_{\rm crit}^{\rm Skyrme}=\sqrt{2}$) and the 
BIS model ($R_{\rm crit}^{\rm BIS}=1.19$) are found as follows:
\begin{eqnarray}
\rho_{B}^{\rm Skyrme}&\equiv& \ltk 2\pi^2 \lk R_{\rm crit}^{\rm Skyrme}\rk^3 \rtk^{-1}
\lk ef_\pi \rk^3 \simeq 4.26 \rho_0, \label{rho_c_SS}\\
\rho_{B}^{\rm BIS}&\equiv& \ltk 2\pi^2 \lk R_{\rm crit}^{\rm BIS}\rk^3 \rtk^{-1}
\lk ef_\pi \rk^3 \simeq 7.12 \rho_0. \label{rho_c_BIS}
\end{eqnarray}
The heavy 
$\rho$ meson fields in the core region of the 
baryon tend to
decrease the total size of the baryon~\cite{NSK},
so that the $\rho$ meson field has 
a significant role to increase the critical density
into the uniform phase as in Eq.~(\ref{rho_c_BIS})
relative to Eq.~(\ref{rho_c_SS}).

In Sec.~\ref{ECpro},
we discuss the disappearance of the $\rho$ meson field
in high density phase of baryonic matter,
while we now find the significant roles of
$\rho$ meson field as for the critical density
in Eq.~(\ref{rho_c_BIS})
relative to Eq.~(\ref{rho_c_SS}).
This situation is somehow resemble to the
``two-Higgs model'' with scalar fields $\phi$ and $\chi$
in the finite temperature.
On the critical temperature of
a phase transition with the dynamics of a scalar filed $\chi$,
the condensate of the $\chi$ field
might become trivial as $\langle\chi\rangle=0$.
However, in general, the mass of $\chi$ field
and also the interactions between $\chi$ and $\phi$ fields 
affect the critical phenomena like critical temperature.
In this sense, even if $\rho$ meson field disappear
near critical point,
they could affect the critical phenomena 
as in Eq.~(\ref{rho_c_BIS})
through its mass and also the interactions with pions
in the action $S_{\rm eff}$ in Eq.~(\ref{DBI}).

Here we briefly comment about the value of the critical density 
by taking other parameter set used by Adkins {\it et~al.}~\cite{ANW}.
In 1983,
Adkins {\it et~al.} 
analyzed the baryon mass spectra with the semiclassical quantization
of the Skyrme model,
whereas the $N$-$\Delta$ splitting corresponds to the higher order
contribution $O(N_c^{-1})$ relative to the static Skyrme action
$O(N_c^{1})$
in the large-$N_c$ expansion.
To reproduce the proper amount of the $N$-$\Delta$ splitting,
they take the smaller pion decay constant and Skyrme parameter~\cite{ANW} as
\begin{eqnarray}
f_\pi=64.5 {\rm MeV},\hspace{4mm} e=5.44. \label{ANW_input1}
\end{eqnarray}
Now if the parameter set in Eq.(\ref{ANW_input1}) is used for the
baryon-number density in Eq.~(\ref{rho_cri1}),
we find too small critical density $(\sim 0.6\rho_0)$
for the Skyrme model.
If one can relate the phase transition in the Skyrme model
with the deconfinement of the baryon in QCD,
it might propose, e.g.,
that quark degrees of freedom are
already manifest in the normal nuclei.  
This unreasonable results may come from the fact that
semiclassical quantization procedure corresponds to
extracting higher order contributions $O(N_c^{-1})$ from the leading order 
Skyrme action $O(N_c^{1})$
as for the large-$N_c$ expansion.

Same comments can also be 
applied to
the holographic model
if one try to perform the semiclassical quantization as the baryon analysis.
In fact,
holographic QCD is derived as the large-$N_c$ effective theory
even by starting from superstring framework.
Higher order effects of the large-$N_c$ expansion 
like the baryon mass splitting
correspond to
the string loop effects beyond the classical supergravity,
which should be fairly untractable.
With these considerations above, we do not intentionally 
proceed to the semiclassical
quantization of Skyrme soliton
in the present work,
and we employ the experimental inputs 
(\ref{exp_input1}) and Skyrme parameter (\ref{e_sky_exp_input})
in the analysis of the critical densities 
with recovering the physical units.
Actually,
due to the scaling property of the BIS model discussed in
Sec.~\ref{MEA},
all the results in ANW unit in the previous sections 
should not be altered,
being independent of the definite values for $f_\pi$ and $m_\rho$.

\section{Summary and outlook \label{SO}}
We have studied baryonic matter in holographic QCD
with ${\rm D}4/{\rm D}8/\overline{{\rm D}8}$ multi-D brane configurations,
by analyzing the system of single brane-induced Skyrmion
on the three-dimensional closed manifold $S^3$.
By changing the size of $S^3$,
the density dependence of the baryon properties 
are examined from a holographic point of view.

First 
we begin with the 
Dirac-Born-Infeld (DBI) action of the probe D8 brane
with D4 supergravity background.
With the dimensional reductions,
we get the five-dimensional Yang-Mills action with a curved
fifth dimension, 
as the leading order of the large-$N_c$ and large 'tHooft coupling expansions
in dual of non-perturbative (strong-coupling) QCD.
Through the mode expansions of the five-dimensional gauge fields,
we get the four-dimensional meson effective action
from holographic QCD.
Especially,
we emphasize the appearance
of the ultraviolet cutoff scale
$M_{\rm KK}\sim 1{\rm GeV}$ in the holographic approach
to be dual of QCD.
Therefore we construct the four-dimensional 
meson effective action
with pion and $\rho$ meson fields below
the cutoff scale $M_{\rm KK}$.

Then we discuss the baryon in holographic QCD
as the ``brane-induced Skyrmion''
in the four-dimensional meson effective action.
The analyses
of the baryon
in the 
meson effective action
with restricted degrees of freedom below the cutoff scale 
$M_{\rm KK}$
is called the ``truncated resonance model'' for the baryon.
Taking
the hedgehog configuration Ansatz for pion and $\rho$ meson fields,
%
we can investigate many properties of single baryon as
the brane-induced Skyrmion,
which are inclusively summarized in Ref.~\cite{NSK}.

Then
we consider the baryonic matter in holographic QCD, as the extension of
holographic approach to dense QCD.
Especially we treat the baryonic matter with large-$N_c$,
because holographic QCD is 
derived as the large-$N_c$ effective theory,
in dual of the classical supergravity. 
For sufficiently large $N_c$,
the kinetic energy and the quantum effects can be 
suppressed relative to the static mass
from simple large-$N_c$ countings~\cite{tH},
where the baryonic matter comes into the static Skyrme matter.
%
In this sense,
we analyze the static Skyrme matter
to see the 
typical features of the baryonic matter with large-$N_c$ conditions.

In order to analyze the static Skyrme matter on the flat coordinate
space ${\boldmath \mbox{${\rm R}$}}^3$,
we alternately treat the system of single brane-induced Skyrmion
on the three-dimensional closed manifold $S^3$.
The interactions between the baryons are simulated by the curvature of the closed manifold $S^3$,
and,
as the size of $S^3$ decreases, the
baryon-number density
increases in this modeling.
Actually, through
the projection procedure from the flat space 
${\boldmath \mbox{${\rm R}$}}^3$ onto the curved space $S^3$,
we get the hedgehog mass 
and also the Euler-Lagrange equations for
pion and $\rho$ meson fields of the brane-induced Skyrmion on $S^3$.

By numerically solving the
Euler-Lagrange equations 
for pion and $\rho$ meson fields on $S^3$,
we find a stable soliton solution as the brane-induced Skyrmion
on $S^3$.
By using this solution,
we analyze many properties of the baryon within the baryonic matter.
Especially by comparing the standard Skyrme model without $\rho$ mesons and
the brane-induced Skyrme (BIS) model,
the roles of (axial) vector mesons in dense QCD are discussed
from a holographic point of view.

First we show the baryon-number density dependence of 
the energy density and the field configuration profiles of single baryon, 
by changing the size of $S^3$.
As the size of $S^3$ decreases,
the localized energy density distribution of single baryon
becomes uniform one as the identity map, 
which is called the ``delocalization phase transition''.
The critical radii of the phase transitions 
are given as $R=R_{\rm crit}^{\rm Skyrme}=\sqrt{2} [\frac{1}{ef_\pi}]$  for the Skyrme model 
and $R=R_{\rm crit}^{\rm BIS}=1.19 [\frac{1}{ef_\pi}]$ for the BIS model.
%
%
Because of the shrinkage of the baryon size 
due to the $\rho$ meson effects~\cite{NSK},
the smaller critical radius of $S^3$, i.e.,
the larger baryon-number density
is needed for the BIS model to give the delocalization phase transition.
We also find
that $\rho$ meson field absolutely disappears
and only pion field survives 
near the critical density.
Then,
with the mathematical arguments,
we propose a remarkable conjecture that
all the (axial) vector meson fields would disappear and only pion field survives
near the critical density, referred  
as the ``pion dominance'' 
in dense baryonic matter.

We also investigated
the baryon-number density dependence of the mass and the root-mean-square (RMS)
mass radius of single baryon.
%
We find the the ``swelling'' phenomenon of the baryon,
as the non-linear increase of the RMS mass radius near the critical density.
%
%
Actually,
non-perturbative QCD vacuum around the baryons are 
closely packed as the baryon-number density increases.
Therefore,
the non-perturbative effect with long wave length like the bag pressure 
would be blocked out from such
definite region to give the swelling of the baryon.
Such swelling provides the the decrease of the baryon mass splitting,
indicating the lost of the individualities for the baryons
as the precursor of the deconfinement.
%
%
We also propose the stable
$N$-$\Delta$ mixed matter in dense QCD because of the swelling,
which could, e.g., soften the EOS of the neutron stars.

The features of the delocalization phase transition and the
chiral symmetry restoration 
are also analyzed with the order parameters in the holographic QCD.
We conjecture with careful arguments
that
the delocalization phase transition can be related with
the deconfinement of the baryons in QCD.
Then we find the coherence of the deconfinement and
the chiral symmetry restoration
through the meson field configurations.
Such coherence of the two phase transitions
are also suggested in the lattice QCD study in the finite temperature~\cite{Rothe},
the PNJL model as a low-energy effective theory of QCD~\cite{Weise},
and also the Sakai-Sugimoto model
as one of the reliable holographic approaches~\cite{NSK}.

We also calculate the critical densities of the phase transitions
with the experimental inputs for the pion decay constant $f_\pi$
and $\rho$ meson mass $m_\rho$ as
$f_\pi=92.4 {\rm MeV}$ and $m_\rho=776.0 {\rm MeV}$.
We find the critical densities as $\rho_B^{\rm Skyrme}\simeq 4.26 \rho_0$ 
for the Skyrme model and 
$\rho_B^{\rm BIS}\simeq 7.12 \rho_0$
for the BIS model.
%
We can see that the larger baryon-number density is needed
for the BIS model to give the phase transitions,
because of the shrinkage of total size of the baryon 
with heavy $\rho$ mesons in its core region.


Finally we compare
our truncated resonance approach for the baryons
with
the other works of baryons as the {\it instantons} in holographic QCD.
In fact, there seem to exist some conflicts for the baryon analysis
in holographic QCD, especially between the truncated resonance model 
and 
the instanton models.
In Refs.~\cite{HSSY,HRYY,HSS,HMY},
the baryons are studied as the instantons on the 
five-dimensional gauge theory of the probe D8 brane with D4 supergravity
background as the holographic dual of QCD.
The properties of the baryonic matter are also analyzed by the system of the single instanton
on the three-dimensional closed manifold $S^3$ as the Wigner-Seiz approximation~\cite{d_Inst}.
In these analyses,
the instanton is found to shrink into zero size 
only for the
DBI sector of the effective action of the 
D8 brane,
which corresponds to the leading order of large-$N_c$ and
large 'tHooft coupling expansions.
In the holographic approach, 
infinite tower of color-singlet 
modes with mesonic quantum numbers as 
$\rho$, $a_1$, $\rho'$, $a_1'$, $\rho''$, $\cdots$,
appears in the mode expansions 
of the five-dimensional gauge field $A_\mu (x,z)$ on the D8 brane as in Eq.(\ref{limit2_lr}).
In this sense, the baryon as the instanton on the five-dimensional gauge theory
before
the mode expansion is in principle to be composed
by the infinite tower of such color-singlet modes with mesonic quantum numbers.
Therefore, the instability of the instanton only with the DBI sector
is often regarded that
the infinite number of (axial) vector mesons 
$\rho$, $a_1$, $\rho'$, $a_1'$, $\rho''$, $\cdots$,
would
affect the low-energy soliton feature to give the zero size.
Then,
in order to describe the baryon
as an instanton with finite size
in the holographic approach,
the inclusion of the Chern-Simons (CS) sector of the effective action of the D8 brane is claimed,
whereas the CS sector corresponds to the higher order contribution of 
the 'tHooft coupling expansion relative to the DBI sector.

However, 
as emphatically 
noted in Sec.~\ref{MEA}, 
the appearance of certain cutoff scale
like $M_{\rm KK}\sim 1{\rm GeV}$ in holographic approach
should be essential to be dual of QCD.
In fact, 
the mesonic mass spectra predicted from holographic QCD
starts to deviate
from the experimental data
beyond the $M_{\rm KK}$ scale,
indicating that the holographic duality with QCD
is mainly maintained almost below $M_{\rm KK}$.
This tendency of the mass spectra
also denotes that such color-singlet modes
beyond cutoff scale $M_{\rm KK}$
do not directly correspond to physical mesons in QCD.
Therefore, one {\it cannot} manifestly conclude that the instability 
of the instanton
only with the DBI sector
really arises from
the physical effect in QCD.

Furthermore,
there also exist the infinite number of non-QCD modes, e.g., the Kaluza-Klein modes 
with large mass $\sim O(M_{\rm KK})$
through 
the Kaluza-Klein compactification of the
D4 brane~\cite{SS}.
%
%
Therefore, if the baryon as the instanton is to be really composed by the infinite tower of 
the color-singlet non-QCD modes even beyond $M_{\rm KK}$,
there is no reason to cast away the Kaluza-Klein modes in the baryon analysis.
In fact, such Kaluza-Klein modes still have the possibility to affect baryon properties,
giving, e.g., the stability of the instanton even without the CS sector of the effective action of the D8 brane,
through it is not to be manifestly proved yet.
As a whole,
there still seems to remain puzzling conflicts between the truncated 
resonance model and 
instanton models for the baryon analyses in holographic QCD.

By looking back the long history of hadron physics,
baryons seem to have inherent difficulties to be described relative to mesons.
%
%
For example, 
the quark model succeeded in
the systematic classification of the hadrons
in terms of their valence quarks~\cite{Gell,Zweig}.
%
However, while the dynamics of mesons as the two-body composites
can be relatively easily analyzed by solving the 
Bethe-Salpeter equation~\cite{BS},
we have to treat the complicated Faddeev equation~\cite{FE}
to describe the dynamics of the baryons 
as the three-body composites.
Same difficulties are also found in the lattice QCD 
studies,
for instance, with respect to the inter-quark potential of mesons~\cite{meson_P}
and baryons~\cite{baryon_P}.
%
%
In fact,
the quark-antiquark potential as the internal nature of the meson
could be successfully measured in lattice QCD in 1980~\cite{q_bar_q}, 
while it takes almost twenty years after that to find the
well measurement of three-quark potential
with the appearance of gluonic Y-type flux tube
as the internal nature of the baryon~\cite{Takahashi}.
%
%
The difficulty of the baryons gets more apparent in 
the large-$N_c$ QCD~\cite{tH},
which has
provided a powerful perturbative treatment
with the $1/N_c$ expansion for
the non-perturbative aspects of the strong interaction.
In fact,
large-$N_c$ QCD achieved large 
successes in the explanations for 
the hadron phenomenology, e.g., the 
Okubo-Zweig-Iizuka (OZI) rule \cite{Witten}, the $\Delta I=1/2$ rule \cite{I=1/2}, 
and the narrowness of meson resonances
\cite{tH, Witten}.
It also provided a quantitative formula for the large
$\eta'$ meson mass as the Veneziano-Witten formula \cite{Veneziano}.
On the other hand,
the baryon does not even appear as the dynamical degrees of freedom
in the large-$N_c$ QCD, because the baryon mass is proportional 
to $O(N_c)$. 
In other words,
$N_c$ quarks are needed to construct the SU($N_c$) color-singlet composite
as a baryon only from quarks,
while a meson can still be constructed as the two-body color-singlet composite
from a quark and an antiquark.
In the end, the baryon appears as a ``non-local object'', i.e.,
the soliton of the meson fields in the large-$N_c$ QCD.
As a whole,
the baryons tend to suffer from the many-body difficulties relative to the mesons,
by seeing the long hadron history.

Now, we suppose here that the recent 
baryon analyses 
in holographic QCD may suffer from the same 
kind of difficulties inherent in
the baryon itself.
%
%
In fact, 
meson properties are successfully described in the framework of holographic QCD as the 
Sakai-Sugimoto model~\cite{SS}
whereas the baryons seem to be not sufficiently described yet.
One should note that the baryon mass splitting corresponds to the higher-order contributions
of the large-$N_c$ expansion.
In terms of the superstring theory, such baryon mass splitting
corresponds to the string loop effect beyond the classical supergravity,
which is fairly untractable.
These naive considerations imply that
the baryon still drags the essential difficulties
even 
by starting from the superstring theory.
We have to sincerely reconsider 
the meaning of the difficulty existing in the baryon itself
and identify an origin of the problems,
which would 
inspire us 
to cure the recent conflicts of 
the baryon analyses
in the holographic approach
for the future.
%

\section*{Acknowledgements}
Authors thank Josh Erlich and Dong-Pil Min for their communications
about our truncated resonance model as the baryon analysis in holographic QCD
in comparison with the instanton models.
K.N. is also indebted to Atsushi Hosaka and Hiroshi Toki
for their discussions
in Research Center for Nuclear Physics (RCNP).
H.S. is supported by a Grant-in-Aid for Scientific Research
[(C) No. 19540287] in Japan.
T.K. is supported by Special Postdoctoral Research Program of RIKEN.
This work is supported by the Global COE Program,
``The Next Generation of Physics, Spun from Universality and Emergence''.



\end{document}